\documentclass[twocolumn]{autart}    
\usepackage{amsfonts}
\usepackage{xcolor}
\usepackage{graphicx}          
\usepackage{epstopdf}
\usepackage{amsmath}
\usepackage{amssymb}
\usepackage{bm}
\usepackage[round]{natbib}
\usepackage{algpseudocode,algorithm}
\usepackage{multirow}
\usepackage{subfigure}
\usepackage{enumerate}
\usepackage{textcomp}
\usepackage{mathrsfs} 
\usepackage{setspace}
\usepackage{nomencl}
\usepackage{threeparttable}
\usepackage{comment}
\usepackage{appendix}
\usepackage{wrapfig}
\usepackage[colorlinks=true, linkcolor=blue, urlcolor=red, citecolor=black]{hyperref}
\usepackage{lipsum}

\graphicspath{{Graphics/}}
\newtheorem{lemma}{\bf Lemma}
\newtheorem{assumption}{Assumption}
\newtheorem{theorem}{\bf Theorem}
\newtheorem{remark}{\bf Remark}
\newtheorem{corollary}{Corollary}

\begin{document}
	\setlength{\abovedisplayskip}{3pt}
	\setlength{\belowdisplayskip}{3pt}
	\begin{frontmatter}
		
		\title{Bias-Variance Trade-off in Kalman Filter-Based Disturbance Observers} 
		
		\thanks[footnoteinfo]{
			The work by D. Shi is supported by the National Natural Science Foundation of China under Grant 62261160575. The work by S. Li is supported in part by the National Natural Science Foundation of China under Grant 62403053 and in part by the National Foreign Experts Program
			under Grant H20251098. The work by X. Lyu, J. Tang and L. Shi is supported by the Hong Kong RGC General Research Fund 16203723. (Corresponding author: Dawei Shi.)}
		\author[ad1]{Shilei Li}\ead{shileili@bit.edu.cn}, 
		\author[ad1]{Dawei Shi}\ead{daweishi@bit.edu.cn} ,
		\author[ad2]{Xiaoxu Lyu}\ead{eelyuxiaoxu@ust.hk} , 
		\author[ad2]{Jiawei Tang}\ead{jtangas@connect.ust.hk} ,  
		\author[ad2,ad3]{Ling Shi}\ead{eesling@ust.hk}  
		
		\address[ad1]{School of Automation, Beijing Institute of Technology, Beijing 100081, China}  
		\address[ad2]{Department of Electronic and Computer Engineering, Hong Kong University of Science and Technology, Clear Water Bay, Kowloon, Hong Kong,
			China} 
		\address[ad3]{Department of Chemical and Biological Engineering, Hong Kong University of Science and Technology, Clear Water Bay, Kowloon, Hong Kong,
			China} 
		
		
		\begin{abstract}                         
			The performance of disturbance observers is strongly influenced by the level of prior knowledge about the disturbance model. The simultaneous input and state estimation (SISE) algorithm is widely recognized for providing unbiased minimum-variance estimates under arbitrary disturbance models. In contrast, the Kalman filter-based disturbance observer (KF-DOB) achieves minimum mean-square error estimation when the disturbance model is fully specified. However, practical scenarios often fall between these extremes, where only partial knowledge of the disturbance model is available. This paper investigates the inherent bias-variance trade-off in KF-DOB when the disturbance model is incomplete. We reveal that SISE can be interpreted as a special case of KF-DOB, where the disturbance noise covariance tends to infinity. To address this trade-off, we propose two novel estimators: the multi-kernel correntropy Kalman filter-based disturbance observer (MKCKF-DOB) and the interacting multiple models Kalman filter-based disturbance observer (IMMKF-DOB). Simulations verify the effectiveness of the proposed methods.
		\end{abstract}
	\end{frontmatter}
	
	\section{Introduction}
	Disturbances are pervasive across various domains, including cyber-physical systems~\citep{b4}, robotics~\citep{b1}, and physiological systems~\citep{bb4}. In control systems, disturbances degrade system performance and can even lead to instability~\citep{b5}. In estimation systems, disturbances diminish estimation accuracy and may cause filter divergence~\citep{bb5}.  Therefore, accurately estimating the disturbance is crucial for a wide range of applications. 
	
	A promising approach to estimate disturbance is disturbance observers (DOBs). DOBs can be roughly divided into two categories: frequency domain-based methods and time domain-based methods. The frequency domain-based approach was initially developed by ~\citet{b6} where the inverse of the nominal model $G(s)$ accompanied by a low-pass filter $Q(s)$ was utilized to estimate the disturbance. To improve the convergence rate with periodic disturbances, \citet{b8} enhanced the conventional frequency-domain disturbance observer by mixing it with a series of multi-resonant terms. The time domain-based DOBs are much more pervasive than the frequency-based approaches, e.g., nonlinear disturbance observer (NDOB)~\citep{b9}, higher-order NDOB~\citep{b10}, extended state-observer (ESO)~\citep{b11}, unknown input observer (UIO)~\citep{b12}, simultaneous input and state estimation (SISE)~\citep{b13,b19}, Kalman filter-based disturbance observer (KF-DOB)~\citep{b14}, to name only a few. \citet{b9} designed an NDOB to offset the friction torque in manipulators with guaranteed stability. \citet{b10} proposed a generalized form of NDOB by considering higher-order disturbances. Different from NDOB, ~\citet{b11} designed ESO by utilizing the relative degree of the system information and estimating the lumped disturbance and state simultaneously. This method is widely used in active disturbance rejection control (ADRC). Opposite to the above estimators that do not require an explicit disturbance model, UIO assumes that the disturbance is generated by an exogenous system with a known differential equation. Then, the state and disturbance can be jointly estimated~\citep{b12}. It is worth mentioning that NDOB, ESO, and UIO do not consider measurement and process noise explicitly. However, in some applications, the noise cannot be ignored. In such a scenario, KF-DOB is preferable.

	The state estimation of linear systems with unknown disturbance (or input) and stochastic measurements has attracted many research efforts since its inception. Initially, \citet{b17} formulated an unbiased minimum-variance estimator for robust state estimation under arbitrary disturbances. Afterward, ~\citet{b18} developed an alternative unbiased minimum variance estimator and provided the corresponding convergence analysis. Subsequently, ~\citet{b19} refined the results of ~\citet{b17} and constructed the SISE estimator which provided an unbiased minimum-variance estimation for both the state and disturbance. Recently, \citet{b20} interpreted SISE as a special KF with a specific disturbance model and \citet{b21} further extended this result to the direct feedthrough case. It is worth noting that although the above estimators are theoretically attractive due to the unbiased minimum variance property, they are rarely utilized in practical control engineering since their outputs usually contain a high uncertainty (in other words, the outputs are noisy). The noisy estimates are detrimental to the control stability and energy efficiency of robots. On the contrary, the result of KF-DOB is much more smooth (although it may be biased) and is favored by many engineers~\citep{b15_1}.
	
	Intuitively, the disturbance estimation is idealistic if the result is both unbiased and smooth. However, in the presence of incomplete disturbance models and stochastic measurements, achieving these two objectives is inherently conflicting. We refer to this effect as the disturbance estimation bias-variance dilemma. This phenomenon has been frequently observed by many researchers. For instance, in KF-DOB, reducing the disturbance noise covariance $Q_d$ results in smoother output but compromises tracking speed. Although the trade-off between tracking speed and estimation smoothness is well understood empirically, an in-depth analysis of this trade-off within the KF framework, along with effective countermeasures, remains unexplored, which motivates this work.
	
	This paper uncovers the inherent bias-variance trade-off in KF-DOB and proposes corresponding solutions. First, we demonstrate the existence of this trade-off in KF-DOB under inaccurate disturbance models and stochastic noise. We then establish the equivalence between SISE and KF-DOB when the disturbance noise covariance is infinite, highlighting the critical role of disturbance noise covariance in balancing the bias-variance effect. Note that the primary connection between SISE and KF was built in \citet{b20}. In this work, we extend the result of \citet{b20} to the scenario of knowing nominal disturbance models, which are commonly utilized in the robotic community ~\citep{b15,b15_1}. Finally, we propose two remedies to mitigate the bias-variance dilemma in KF-DOB, i.e., MKCKF-DOB and IMMKF-DOB. The main contributions of this work are outlined as follows.     
	\begin{itemize}
		\item We establish the existence of a bias-variance trade-off in KF-DOB, as formalized in \textbf{Theorem} \ref{theorem3}. Additionally, we prove that KF-DOB is equivalent to SISE when the disturbance noise covariance tends to infinity, as demonstrated in \textbf{Theorem} \ref{theorem5} and \textbf{Corollary} \ref{corollary3}.
		\item We develop two remedies to alleviate the bias-variance trade-off in KF-DOB, i.e., MKCKF-DOB and IMMKF-DOB in \textbf{Algorithms \ref{mkckf-dob}  and \ref{immkf-dob}}. 
		\item The bias-variance effects of different estimators are visualized in simulations. The results confirm that the proposed methods offer superior performance compared to existing estimators.
	\end{itemize}
	
	The remainder of this paper is arranged as follows. In Section II, we provide some preliminaries. In Section III, we demonstrate the intrinsic bias-variance trade-off in KF-DOB. In Section IV, we prove that KF-DOB is identical to SISE when applying infinite disturbance noise covariance and develop two remedies. In Section V, we give some simulations to validate the proposed approaches. In Section VI, we draw a conclusion. 
	
	\emph{Notations}: The symbol $X\succ 0$ ($X \succeq 0$) denotes $X$ is a positive definite (semi-positive definite) matrix. The spectral norm of a matrix $X$ is denoted as $\|X\|$. The Gaussian distribution with mean $\mu$ and covariance $\Sigma$ is denoted by $\mathcal{N}(\mu,\Sigma)$. The expectation of a random variable $X$ is denoted by $E(X)$. The notation $A \to \infty$ denotes that all eigenvalues of $A$ tend to infinity, which implies that $A^{-1} \to 0$. The symbol $\|x\|_{A}^{2}$ denotes $x^{T}Ax$. 
	
	\section{Preliminaries}
	We begin by revisiting the SISE estimator and KF-DOB. Next, we present the multi-kernel correntropy Kalman filter (MKCKF) and the interacting multiple models Kalman filter (IMMKF). Finally, we summarize the distinct characteristics of each estimator and provide a problem description.
	\subsection{SISE Algorithm}
	We consider the following linear system:
	\begin{equation}
		\begin{aligned}
			{x}_{k}&={F}_k {x}_{k-1}+{G}_k {d}_{k-1}+{w}_{k}\\
			{y}_k&={H}_k {x}_{k}+{v}_{k}
			\label{linearfun}
		\end{aligned}
	\end{equation}
	where $x_k \in \mathbb{R}^{n}$ is the state, $y_k \in \mathbb{R}^{m}$ is the measurement, $d_k \in \mathbb{R}^{p}$ is the unknown input, $w_k \sim \mathcal{N}(0, Q_k)$, and $v_k \sim \mathcal{N}(0,R_k)$. The SISE algorithm \citep{b19} is summarized as follows:
	\begin{enumerate}[1)]
		\item time update
		\begin{equation}
			\begin{aligned}
				x_{k|k-1}&=F_k x_{k-1|k-1}\\
				P_{k|k-1}&=F_k P_{k-1|k-1} F_k^{T}+Q_k.
				\label{sise1}
			\end{aligned}
		\end{equation} 
		\item unknown input estimation
		\begin{equation}
			\begin{aligned}
				\tilde{R}_k&=H_k P_{k|k-1} H_k^{T} + R_k \\
				M_k^{*} & =\left(G_{k}^T H_k^T \tilde{R}_k^{-1} H_k G_{k}\right)^{-1} G_{k}^T H_k^T \widetilde{R}_k^{-1}\\
				{d}_{k-1} & =M_k^{*}\left(y_k-H_k {x}_{k|k-1}\right).
				\label{sise2}
			\end{aligned}
		\end{equation}
		\item measurement update
		\begin{equation}
			\scriptsize		
			\begin{aligned}
				{x}_{k \mid k}^*= & {x}_{k \mid k-1}+G_{k} {d}_{k-1} \\
				K_k^{*}= & P_{k \mid k-1} H_k^T \widetilde{R}_k^{-1} \\
				\hat{x}_{k \mid k}= & {x}_{k \mid k}^*+K_k^{*} \left(y_k-H_k {x}_{k \mid k}^*\right) \\
				P_{k \mid k}= & \left(I-K_k^{*} H_k\right)\Big[\left(I-G_{k} M_k H_k\right) P_{k \mid k-1} \left(I-G_{k} M_k H_k\right)^T\\
				&+G_{k} M_k R_k M_k^T G_{k}^T\Big]+K_k^{*} R_k M_k^T G_{k}^T\\
				P_{k|k}^{dd}=&\left(G_{k}^T H_k^T \tilde{R}_k^{-1} H_k G_{k}\right)^{-1}.
				\label{sise3}
			\end{aligned}
		\end{equation}
	\end{enumerate}
	\begin{lemma}\citep{b19}
		$M_k^{*} H_k G_k = I_{p}$ where $I_{p}$ is an identity matrix of dimension $p$.
		\label{lemma1}
	\end{lemma}
	\begin{lemma}\citep{b19}
		The SISE algorithm is an unbiased minimum variance estimator (UMVE) under arbitrary disturbance signals. 
		\label{lemma2}
	\end{lemma}
	\subsection{Kalman Filter-Based Disturbance Observer}
	In many applications, one can augment the disturbance as a new state element and  construct the KF-DOB as follows:
	\begin{equation}
		\begin{aligned}
			\mathrm{x}_k &= \Phi_k \mathrm{x}_{k-1} + \mathrm{w}_k \\
			\mathrm{y}_k &= \mathrm{H}_k \mathrm{x}_k  + \mathrm{v}_k
		\end{aligned}
		\label{sys}
	\end{equation}
	where 
	\begin{equation}\nonumber
		\begin{aligned}
			\mathrm{x}_{k}=\begin{bmatrix}
				d_{k} \\
				x_{k}
			\end{bmatrix}, \Phi_k=\begin{bmatrix}
				I   &0\\
				G_k &F_k 
			\end{bmatrix}, \mathrm{H}_k =\begin{bmatrix}
				{0} & H_k
			\end{bmatrix}, \mathrm{y}_k = y_k.
		\end{aligned}
		\label{notdef}
	\end{equation}
	The augmented process noise is assumed to be distributed as $\mathrm{w}_k \triangleq [w_{d,k}^{T}, w_{x,k}^{T}]^{T} \sim \mathcal{N}( 0, \mathrm{Q}_k)$ and $\mathrm{v}_k \triangleq v_k \sim \mathcal{N}(0, \mathrm{R}_k)$. Correspondingly, the KF-DOB can be executed as
	\begin{equation}
		\begin{aligned}
			\mathrm{\hat{x}}_{k|k-1} &= \Phi_k \mathrm{\hat{x}}_{k-1|k-1}\\
			\mathrm{\hat{x}}_{k|k} &= \mathrm{\hat{x}}_{k|k-1}+\mathrm{K}_k(\mathrm{y}_k-\mathrm{H}_k \mathrm{x}_{k|k-1}) \\
			&=(\mathrm{I}-\mathrm{K}_k \mathrm{H}_k)\Phi_k \mathrm{\hat{x}}_{k-1|k-1} + \mathrm{K}_k \mathrm{y}_k \\
			\mathrm{P}_{k|k-1} &= \Phi_k \mathrm{P}_{k-1|k-1} \Phi_k^{T}  +  \mathrm{Q}_k\\
			\mathrm{K}_k &= \mathrm{P}_{k|k-1} \mathrm{H}_k^{T} (\mathrm{H}_k \mathrm{P}_{k|k-1} \mathrm{H}_k^{T}+\mathrm{R}_{k})^{-1}\\
			\mathrm{P}_{k|k}&=(\mathrm{I}-\mathrm{K}_k \mathrm{H}_k )\mathrm{P}_{k|k-1}(\mathrm{I}-\mathrm{K}_k \mathrm{H}_k )^{T}+\mathrm{K}_k \mathrm{R}_{k} \mathrm{K}_k^{T}.
			\label{pk}
		\end{aligned}
	\end{equation}
	The information form posterior error covariance update~\citep{c7} has
	\begin{equation}
		\begin{aligned}
			\mathrm{P}_{k|k}^{-1}=\mathrm{P}_{k|k-1}^{-1} + \mathrm{H}_k^{T} \mathrm{R}_k^{-1} \mathrm{H}_k.
			\label{infp}
		\end{aligned}
	\end{equation}
	\begin{lemma}
		For the Kalman filter shown in \eqref{pk}, by denoting $\mathrm{M}_k=\mathrm{I}-\mathrm{K}_k \mathrm{H}_k$, one has
		\begin{equation}
			\begin{aligned}
				\mathrm{M}_k=\big(\mathrm{I}+ \mathrm{P}_{k|k-1} \mathrm{H}_k^{T} \mathrm{R}_k^{-1} \mathrm{H}_k \big)^{-1}.
			\end{aligned}
			\label{ident}
		\end{equation}
		\label{lemma5}
	\end{lemma}
	The proof is available in Appendix \ref{appendix4}. The difference between the SISE and KF-DOB is that a nominal dynamic model $d_k=d_{k-1} + w_{d,k}$ is utilized in KF-DOB, but is avoided in SISE. It is worth mentioning that the constant disturbance model can be replaced by other models that reflect the prior knowledge of the disturbance in practical applications.
	\subsection{Multi-kernel Correntropy Kalman Filter}
	The MKC ~\citep{b26,b27} is a similarity measure of two random vectors $\mathscr{X}, \mathscr{Y} \in \mathbb{R}^{l}$: 
	\begin{equation}
		V(\mathscr{X},\mathscr{Y})= \sum_{i=1}^{l} \sigma_i^2 E[ \kappa_{\sigma_i}(\mathscr{X}_i,\mathscr{Y}_i)]
	\end{equation}
	where $E[\kappa_{\sigma_i}(\mathscr{X}_i,\mathscr{Y}_i)]=\int \kappa_{\sigma_i}(x_i,y_i)d F_{\mathscr{X}_i\mathscr{Y}_i}(x_i,y_i)$, $\kappa_{\sigma_i}(x_i,y_i)=G_{\sigma_i}(x_i,y_i)=\exp(-\frac{e_i^2}{2\sigma_i^2})$, $\sigma_i$ is the kernel bandwidth, $e_i=x_i-y_i$ is the realization error, and $F_{\mathscr{X}_i\mathscr{Y}_i}(\cdot,\cdot)$ is the joint distribution. In some applications, only finite samples $x_{k}$ and $y_{k}$ can be obtained. Then, MKC can be estimated as
	\begin{equation}
		\hat{V}(\mathscr{X},\mathscr{Y})=\sum_{i=1}^{l}\sigma_i^2 \hat{V}_{i}(\mathscr{X}_i,\mathscr{Y}_i)
	\end{equation}
	where
	$\hat{V}_{i}(\mathscr{X}_i,\mathscr{Y}_i)= \frac{1}{N}\sum_{k=1}^{N}G_{\sigma_i}\big({x_{i,k},y_{i,k}}\big)$, and $x_{i,k}$ and $y_{i,k}$ denote $i$-th elements of $x_k$ and $y_k$. The MKC loss (MKCL) is defined as 
	\begin{equation}
		\begin{aligned}
			J_{MKCL}&=\sum_{i=1}^{l}\sigma_i^2 (1-\hat{V}_{i})=\frac{1}{N}\sum_{k=1}^{N}\sum_{i=1}^{l}\sigma_i^2 \Big(1-G_{\sigma_i}(e_{i,k})\Big)
			\label{GL}
		\end{aligned}
	\end{equation}
	where $e_k=[e_{1,k},e_{2,k},\ldots,e_{l,k}]^{T}$ and $e_{i,k}=x_{i,k}-y_{i,k}$. 
	\subsection{Interacting Multiple Models Kalman Filter}
	\label{immkf}
	We consider a state estimation problem with $q$ Markov jump linear models. For model $j$ ($j=1,~2,~\ldots,~q$), we have 
	\begin{equation}
		\begin{aligned}
			{x}_{k}&={A}_{j}{x}_{k-1}+{w}_{j,k}\\
			{y}_k&={C}_{j}{x}_k+{v}_{j,k}
			\label{linear}
		\end{aligned}
	\end{equation}
	where $x_k \in \mathbb{R}^{n}$ is the state, $y_{k} \in \mathbb{R}^{m}$ is the measurement,  and ${w}_{j,k}$ and ${v}_{j,k}$ are Gaussian noises for $j$-th model with covariance matrices $Q_{j}$ and $R_{j}$, respectively. The Markov transition probability matrix has $\mathcal{P}=[\mathcal{P}_{i,j}]$ where $\mathcal{P}_{i,j}$ is the transition probability from model $i$ to model $j$. Then, the IMM-KF can be summarized as follows~\citep{c9}:
	\begin{enumerate}[1)]
		\item Obtain the transition probability from model $i$ to model $j$ at time step $k-1$:
		\begin{equation}
			\mu_{ij,k-1|k-1}=\mathcal{P}_{ij}\mu_{i,k-1}/\bar{c}_{j}
			\label{imm1}
		\end{equation}
		where $\mu_{i,k-1}$ is the probability for model $i$ obtained at time step $k-1$ and $\bar{c}_{j}=\sum_{i=1}^{q}\mathcal{P}_{ij}\mu_{i,k-1}$.
		\item  Obtain the initial state and covariance estimate of model $j$:
		\begin{equation}
			\begin{aligned}
				\hat{x}_{j,k-1|k-1}^{init}&=\sum_{i=1}^{q}\hat{x}_{i,k-1|k-1}\mu_{ij,k-1|k-1}\\
				P_{j,k-1|k-1}^{init}&=\sum_{i=1}^{q}\mu_{ij,k-1|k-1}\Big{(}P_{i,k-1|k-1}+\big{(}\hat{x}_{i,k-1|k-1}
				\\&-\hat{x}_{i,k-1|k-1}^{init}\big{)}
				\big{(}\hat{x}_{i,k-1|k-1}-\hat{x}_{i,k-1|k-1}^{init}\big{)}^{T}\Big{)}.
			\end{aligned}
			\label{imm2}
		\end{equation}
		\item State and covariance estimation for model $j$:
		\begin{equation}
			\begin{aligned}
				\hat{x}_{j,k|k}&=\hat{x}_{j,k|k-1}+K_{j,k}e_{j,k}\\
				P_{j,k|k}&=\big{(}I-K_{j,k}C_{j}\big{)}P_{j,k|k-1}
				\label{imm3}
			\end{aligned}
		\end{equation}
		with 
		\begin{equation}
			\begin{aligned}
				\left\{\begin{array}{l}
					\hat{x}_{j,k|k-1}= A_{j}\hat{x}_{j,k-1|k-1}^{init}\\
					P_{j,k|k-1}=A_{j}P_{j,k-1|k-1}^{init}A_{j}^{T}+ Q_{j}\\
					e_{j,k}= y_k-A_{j}\hat{x}_{k|k-1}\\
					S_{j,k}=C_{j}P_{j,k|k-1}C_{j}^{T}+R_{j}\\
					K_{j,k}=P_{j,k|k-1}C_{j}^{T}S_{j,k}^{-1}
				\end{array}\right..
			\end{aligned}
			\label{imm31}
		\end{equation}
		\item Update the model probability for model $j$:
		\begin{equation}
			\mu_{j,k}=\frac{\Lambda_{j,k}\bar{c}_{j}}{c}
			\label{imm4}
		\end{equation}
		where
		\begin{equation}\nonumber
			\begin{aligned}
				\Lambda_{j,k}&=\frac{1}{\sqrt{2\pi|S_{j,k}|}}\exp\big{(}-\frac{1}{2}e_{j,k}^{T}S_{j,k}^{-1}e_{j,k}\big{)}\\
				c &=\sum_{j=1}^{q}\Lambda_{j,k}\bar{c}_{j}.
			\end{aligned}
		\end{equation}
		\item State and error covariance interaction:
		\begin{equation}
			\scriptsize
			\begin{aligned}
				\hat{x}_{k|k}&=\sum_{j=1}^{q}\mu_{j,k}\hat{x}_{j,k|k} \\
				P_{k|k}&=\sum_{j=1}^{q}\mu_{j,k}\Big{(}P_{j,k|k}+\big{(}\hat{x}_{j,k|k}-\hat{x}_{k|k}\big{)} \big{(}\hat{x}_{j,k|k}-\hat{x}_{k|k}\big{)}^{T}\Big{)}.
			\end{aligned}
			\label{imm5}
		\end{equation}
	\end{enumerate}
	\subsection{\textcolor{black}{Problem Description}}
	We summarize the characteristics of the above four estimators in Table \ref{charac}.  We find that SISE is unbiased under arbitrary disturbance models and KF-DOB is optimal under accurate disturbance models. In many practical applications, we only have an inaccurate disturbance model. In such cases, SISE is under-confident since it ignores the disturbance model, and KF-DOB is over-confident since it regards the disturbance model as an accurate one. Since unmodeled disturbance dynamics can be regarded as heavy-tailed noise~\citep{b27} and a linear combination of multiple models can better approximate inaccurate disturbance models, MKCKF and IMMKF have the potential of outperforming SISE and KF-DOB in terms of disturbance estimation.
	\begin{table}[h]
		\centering
		\caption{Characteristics of different estimators.}
		\scalebox{0.75}{
			\begin{tabular}{ccc}
				\hline
				\hline
				Estimators & Noise       & Features                                                             \\
				\hline
				SISE       & Gaussian     & UMVE under \emph{arbitrary disturbance model}    \\
				KF-DOB     & Gaussian     & optimal under \emph{accurate disturbance model} \\
				MKCKF      & heavy-tailed & robust to heavy-tailed noise                                                         \\
				IMMKF      & Gaussian     & capable of Markov jump systems                                     \\
				\hline         
				\hline
		\end{tabular}}
		\label{charac}
	\end{table}
	
	In this paper, we answer the following questions: What is the role of the disturbance model in simultaneous state and disturbance estimation? Can we build a connection between SISE and KF-DOB? How can we provide a better estimator when the disturbance model is inaccurate? The answers would facilitate our understanding of state estimation with inaccurate disturbance models and guide the practical usage of disturbance observers.

	\section{Bias-Variance Trade-off in KF-DOB}
	We investigate the fundamental trade-off between the disturbance tracking speed and estimation variance in KF-DOB. 
	\subsection{Bias-Variance Effects with Incorrect Initialization in KF}
	Directly investigating the bias-variance effect in KF-DOB is difficult. To simplify the problem, we initially investigate the effects of the process covariance selection with an improper initial guess in KF, which would pave the way for the subsequent analysis.
	
	\subsubsection{Effects of $\Delta \mathrm{Q}$ on Convergence Speed}
	We consider the linear time-invariant system as shown in \eqref{sys}. By denoting the used process covariance $\mathrm{Q}_k^{u}$ as
	\begin{equation}
		\mathrm{Q}_k^{u}=\mathrm{Q}_k+ \Delta \mathrm{Q}
		\label{qsys}
	\end{equation}
	where $\mathrm{Q}_k$ denotes the real process noise covariance and $\Delta \mathrm{Q}$ is the covariance mismatch. Subsequently, we investigate the effects of $\Delta \mathrm{Q}$ on the convergence speed and estimation accuracy with incorrect initial guess in KF. 
	
	By aggregating the state and measurement from time step 1 to $k$ as $\mathrm{X}_{1,k}=[\mathrm{x}_{1}^{T},\mathrm{x}_{2}^{T},\cdots, \mathrm{x}_{k}^{T}]^{T}$ and $\mathrm{Y}_{1,k}=[\mathrm{y}_{1}^{T},\mathrm{y}_{2}^{T},\cdots, \mathrm{y}_{k}^{T}]^{T}$, the extended state-space model of \eqref{sys} has
	\begin{equation}
		\begin{aligned}
			\mathrm{X}_{1,k} &= \mathrm{\Phi}_{1,k}\mathrm{x}_0+\mathrm{G}_{1,k}\mathrm{W}_{1,k}\\
			\mathrm{Y}_{1,k} &= \mathrm{H}_{1,k}\mathrm{x}_0+\mathrm{D}_{1,k}\mathrm{W}_{1,k}+ \mathrm{V}_{1,k}\\
		\end{aligned}
		\label{vecsys}
	\end{equation}
	where $\mathrm{x}_0$ denotes the initial state, $\mathrm{W}_{1,k}=[\mathrm{w}_1^{T},\ldots,\mathrm{w}_k^{T}]^{T}$,  $\mathrm{V}_{1,k}=[\mathrm{v}_1^{T},\ldots,\mathrm{v}_k^{T}]^{T}$, and 
	$$
	\begin{aligned}
		\mathrm{\Phi}_{1,k} &= [\Phi_1^{T}, (\Phi_2\Phi_1)^{T}, \ldots, ({\phi}_{k-1}^{1})^{T},({\phi}_{k}^{1})^{T}]^{T}\\
		\mathrm{G}_{1,k}&=\begin{bmatrix}
			\mathrm{I}& 0 &\cdots & 0 &0 \\
			\mathrm{\Phi}_2& \mathrm{I} &\cdots & 0 &0 \\
			\vdots& \vdots &\ddots & \vdots &\vdots \\
			{\phi}_{k-1}^{2}&{\phi}_{k-1}^{3}&\ldots&\mathrm{I}&0\\
			{\phi}_{k}^{2}&{\phi}_{k}^{3}&\ldots&\mathrm{\Phi}_k&\mathrm{I}
		\end{bmatrix}\\
		\mathrm{H}_{1,k} &= \mathrm{\bar{H}}_{1,k} \mathrm{\Phi}_{1,k}
	\end{aligned}
	$$
	where ${\phi}_{j}^{i}=\Phi_i \Phi_{i+1} \ldots\Phi_{j}$ with $i<j$, $\mathrm{\bar{H}}_{1,k}=\operatorname{diag}(\mathrm{H}_1, \mathrm{H}_2,\ldots,\mathrm{H}_k)$, and $\mathrm{D}_{1,k}=\mathrm{\bar{H}}_{1,k}\mathrm{G}_{1,k}$. According to \eqref{vecsys}, we can specify $\mathrm{x}_k$ as the last row vector $\mathrm{X}_{1,k}$: 
	\begin{equation}
		\mathrm{x}_k = \mathrm{\phi}_{k}^{1} \mathrm{x}_0 + 	\mathrm{G}_{1,k}^{rk}\mathrm{W}_{1,k}
		\label{state1}
	\end{equation}
	where $\mathrm{G}_{1,k}^{rk}$ denote $k$-th row of $\mathrm{G}_{1,k}$.
	
	Given the exactly known initial value $\mathrm{x}_0$ and the measurement set $\mathrm{Y}_{1,k} $, the batch Kalman estimate appears in the convolution-based form~\citep{c10} at time step $k$ as 
	\begin{equation}
		\hat{\mathrm{x}}_k = \mathcal{H}_{1, k}^{{h}} \mathrm{Y}_{1,k} + \mathcal{H}_{1, k}^{{s}} \mathrm{x}_{0}
		\label{batest2}
	\end{equation}
	where gains $\mathcal{H}_{1, k}^{{h}}$ and $\mathcal{H}_{1, k}^{{s}}$ minimize the MSE for inputs $\mathrm{Y}_{1,k}$ and $\mathrm{x}_{0}$. According to the unbiasedness constraint and orthogonality principle between the state estimate and the measurement set, one obtains
	\begin{equation}
		\begin{aligned}
			\mathrm{\hat{x}}_k&=\mathrm{\hat{x}}_k^{h}+\mathrm{\hat{x}}_k^{s}=\mathcal{H}_{1, k}^{{h}} \mathrm{Y}_{1,k} + (\mathrm{\phi}_{k}^{1}-\mathcal{H}_{1, k}^{{h}}\mathrm{H}_{1,k}) \mathrm{x}_{0}
		\end{aligned}
		\label{batkf}
	\end{equation}
	where
	\begin{equation}
		\mathcal{H}_{1, k}^{{h}}=\mathrm{G}_{1,k}^{rk} \mathrm{Q}_{1,k}\mathrm{D}_{1,k}^{T}(\mathrm{D}_{1,k}\mathrm{Q}_{1,k}\mathrm{D}_{1,k}^{T}+\mathrm{R}_{1,k})^{-1}.
		\label{gainy}
	\end{equation}
	The detailed derivation is in Appendix \ref{appendix1}. We then consider a much more general case, i.e., $\mathrm{x}_0$ is not exactly known but follows $\mathcal{N}(\bar{\mathrm{x}}_0,\bar{P}_0)$. In this scenario,  the batch estimator \eqref{batest2} is modified as 
	\begin{equation}
		\hat{\mathrm{x}}_k = \mathcal{\bar{H}}_{1, k}^{{h}} \mathrm{Y}_{1,k} + \mathcal{\bar{H}}_{1, k}^{{s}} \mathrm{\bar{x}}_{0}
		\label{batest4}
	\end{equation}
	where $\mathcal{\bar{H}}_{1, k}^{{h}}$ and $\mathcal{\bar{H}}_{1, k}^{{s}}$ are gains to be determined. According to the unbiasedness constraint and orthogonal principle,
	by analogy with the obtainment of \eqref{batkf}, one has 
	\begin{equation}
		\begin{aligned}
			\mathrm{\hat{x}}_k&=\mathrm{\hat{x}}_k^{\bar{h}}+\mathrm{\hat{x}}_k^{\bar{s}}=\mathcal{\bar{H}}_{1, k}^{{h}} \mathrm{Y}_{1,k} + (\mathrm{\phi}_{k}^{1}-\mathcal{\bar{H}}_{1, k}^{{h}}\mathrm{H}_{1,k}) \mathrm{\bar{x}}_{0}
		\end{aligned}
		\label{batkf2}
	\end{equation}
	where
	\begin{equation}
		\begin{aligned}
			\mathcal{\bar{H}}_{1, k}^{{h}}=&(\phi_k^{1}\mathrm{P}_0 \mathrm{H}_{1,k}^{T} +\mathrm{G}_{1,k}^{rk} \mathrm{Q}_{1,k}\mathrm{D}_{1,k}^{T})(\mathrm{H}_{1,k}\mathrm{P}_0 \mathrm{H}_{1,k}^{T}+\\
			&\mathrm{D}_{1,k}\mathrm{Q}_{1,k}\mathrm{D}_{1,k}^{T}+\mathrm{R}_{1,k})^{-1}.
			\label{gainy2}
		\end{aligned}
	\end{equation}
	The derivation is available in Appendix \ref{appendix3}. Since both \eqref{batest4} and the KF in \eqref{pk} are optimal in the minimum mean squared error sense, we have the following lemma.
	\begin{lemma}\citep{c7}
		\label{lemma3}
		Given a linear Gaussian state-space model \eqref{sys} with known initial distribution $\mathcal{N}(\mathrm{\bar{x}}_0,\mathrm{P}_0)$, the optimal state estimate $\mathrm{x}_k$ at time instant $k$ can be obtained either by recursively running conventional KF with $\mathrm{\hat{x}}_0=\mathrm{\bar{x}}_0$ and $\mathrm{P}_{0|0}=\mathrm{P}_{0}$ as shown in \eqref{pk}, or through the batch estimator specified in \eqref{batkf2}.
	\end{lemma}
	\begin{prop}
		The response to measurements ${\mathrm{\hat{x}}_k}^{\bar{h}}$ has
		\begin{equation}
			\mathrm{\hat{x}}_k^{\bar{h}} = (\mathrm{I}-\mathrm{K}_k \mathrm{H}_k)\Phi_k \mathrm{\hat{x}}_{k-1}^{\bar{h}} + \mathrm{K}_k \mathrm{y}_k 
			\label{equ11}
		\end{equation}
		where $\mathrm{P}_{0|0}=\mathrm{P}_{0}$ and $\mathrm{K}_k$ is the Kalman gain. The response to initial value ${\mathrm{\hat{x}}_k}^{\bar{s}}$ has
		\begin{equation}
			\mathrm{\hat{x}}_k^{\bar{s}} = (\mathrm{I}-\mathrm{K}_k \mathrm{H}_k)\Phi_k \mathrm{\hat{x}}_{k-1}^{\bar{s}}
			\label{equ22}
		\end{equation}
		starting from $\mathrm{\hat{x}}_0^{s}=\mathrm{\bar{x}}_0$.
		\label{prop2}
	\end{prop}
	The proof is available in Appendix \ref{addendixprop}. 
	\begin{corollary}
		If KF is stable, one has
		\begin{equation}
			\lim_{k \to \infty}	\mathrm{\hat{x}}_k^{\bar{s}} = \prod_{i=1}^{k}\bar{\Phi}_k \bar{\mathrm{x}}_0^{\bar{s}} =0
		\end{equation} 
		where 
		$\bar{\Phi}_k=(\mathrm{I}-\mathrm{K}_k \mathrm{H}_k)\Phi_k$ and $\bar{\mathrm{x}}_0^{\bar{s}}=\bar{\mathrm{x}}_0$.
		\label{corollary2}
	\end{corollary}
	
	In the case that an incorrect initial mean $\mathrm{\bar{x}}_0^{u}$ is used ( but with correct $\mathrm{P}_{0}$), according to \eqref{equ22}, the corresponding response to initial value has
	$
	\mathrm{\hat{x}}_k^{\bar{s}_u} = \bar{\Phi}_k \mathrm{\hat{x}}_{k-1}^{{\bar{s}_u}}
	$
	where $\mathrm{\hat{x}}_{0}^{{\bar{s}_u}}=\mathrm{\bar{x}}_0^{u}$. Subsequently, the estimation bias can be quantified by  
	\begin{equation}
		\mathrm{\hat{x}}_k^{b}=\bar{\Phi}_k \mathrm{\hat{x}}_{k-1}^{b}.
		\label{biaseffect}
	\end{equation} 
	where $\mathrm{\hat{x}}_k^{b}= \mathrm{\hat{x}}_{k}^{{\bar{s}}} - \mathrm{\hat{x}}_{k}^{{\bar{s}_u}}$ and $\bar{\mathrm{x}}_0^{\mathrm{b}}=\bar{\mathrm{x}}_0-\bar{\mathrm{x}}_0^{u}$. 
	\begin{remark}
		Proposition \ref{prop2} allows us to investigate the estimation bias and estimation uncertainty (i.e., variance), by analyzing \eqref{biaseffect} and \eqref{equ11}, separately. It is worth mentioning that the incorrect initial value can be generalized to the state estimation with sudden state jumps, which can be found in many applications, e.g., target tracking with impulsive disturbances, position tracking of robots with suddenly added loads, etc.
		\label{remark1}
	\end{remark}
	
	Subsequently, we formulate the following convergence performance measure
	\begin{equation}
		\begin{aligned}
			C_{\gamma,k}&=\|\mathrm{\hat{x}}_k^{b}\|_2^{2}=(\mathrm{\hat{x}}_{k-1}^{b})^{T} \bar{\Phi}_k^{T} \bar{\Phi}_k \mathrm{\hat{x}}_{k-1}^{b}\\
			&=(\mathrm{\hat{x}}_{k-1}^{b})^{T} {\Phi}_k^{T}\mathrm{M}_k^{T}\mathrm{M}_k {\Phi}_k  \mathrm{\hat{x}}_{k-1}^{b}
		\end{aligned}
		\label{index}
	\end{equation}
	where $\mathrm{M}_k \triangleq \mathrm{I}-\mathrm{K}_k \mathrm{H}_k$. To investigate effects of $\Delta \mathrm{Q}$ on $C_{\gamma,k}$, we denote $C_{\gamma,k}^{o}$ and $C_{\gamma,k}^{u}$ as the results of applying  $\mathrm{Q}_k$ and  $\mathrm{Q}_k^{u}$. Furthermore,
	we denote
	\begin{equation}\nonumber
		\begin{aligned}
			\mathrm{P}_{k|k-1}^{o}&= \Phi_k \mathrm{P}_{k-1|k-1}^{o} \Phi_k^{T}  +  \mathrm{Q}_k\\
			\mathrm{P}_{k|k-1}^{u}&= \Phi_k \mathrm{P}_{k-1|k-1}^{u} \Phi_k^{T}  +  \mathrm{Q}_k^{u}
		\end{aligned}
	\end{equation} and define auxiliary variables
	\begin{equation}
		\begin{aligned}
			\mathrm{X}_k &\triangleq \mathrm{I}+ \mathrm{P}_{k|k-1}^{o} \mathrm{H}_k^{T} \mathrm{R}_k^{-1} \mathrm{H}_k,\\
			\mathrm{Y}_k &\triangleq \Big(\mathrm{P}_{k|k-1}^{u}-\mathrm{P}_{k|k-1}^{o})\mathrm{H}_k^{T} \mathrm{R}_k^{-1} \mathrm{H}_k.\\
		\end{aligned}
		\label{defXY}
	\end{equation} 
	\begin{assumption}[Special Case]
		\label{assump2}
		The terms $\mathrm{X}_k$ and $\mathrm{Y}_k$ in \eqref{defXY} are symmetric.
	\end{assumption}
	\begin{lemma}[Sufficient Condition]
		Denote
		$\mathbf{Z}_{+} \triangleq (\mathrm{X}_k+\mathrm{Y}_k)^{-1} \mathrm{X}_k$, and $\mathbf{Z}_{-} \triangleq \mathrm{X}_k^{-1}(\mathrm{X}_k+\mathrm{Y}_k)$. 
		Then, the following statements hold:
		\begin{itemize}
			\item If $\|\mathbf{Z}_{+}\| \le 1$, then $(\mathrm{X}_k+\mathrm{Y}_k)(\mathrm{X}_k+\mathrm{Y}_k)^{T}-\mathrm{X}_k \mathrm{X}_k^{T} \succeq 0$.
			\item If $\|\mathbf{Z}_{-}\| \le 1$, then $(\mathrm{X}_k+\mathrm{Y}_k)(\mathrm{X}_k+\mathrm{Y}_k)^{T}-\mathrm{X}_k \mathrm{X}_k^{T} \preceq 0$.
		\end{itemize}
		\label{lemmazz}
	\end{lemma}
	The proof is provided in Appendix \ref{prooflemmazz}.
	\begin{remark} 
		\label{remark2}
		Substituting \eqref{qsys} into \eqref{defXY} yields 
		$\mathrm{Y}_k=\left(\Phi_k (\mathrm{P}_{k-1|k-1}^{u}-\mathrm{P}_{k-1|k-1}^{o}) \Phi_k^{T} +\Delta \mathrm{Q}\right)\mathrm{H}_k^{T} \mathrm{R}_k^{-1} \mathrm{H}_k$. Moreover, we observe that $\mathrm{X}_k$ is not a function of $\Delta \mathrm{Q}$. 
		According to the discrete-time algebraic Riccati equation, $\mathrm{P}_{k-1|k-1}^{u} \succeq \mathrm{P}_{k-1|k-1}^{o}$ if $\Delta \mathrm{Q} \succeq \mathbf{0}$, and $\mathrm{P}_{k-1|k-1}^{u} \preceq \mathrm{P}_{k-1|k-1}^{o}$ if $\Delta \mathrm{Q} \preceq \mathbf{0}$ \citep{sinopoli2004kalman}. 
		This implies that the value of $\|\mathbf{Z}_{+}\|$ or $\|\mathbf{Z}_{-}\|$ can be adjusted by selecting different $\Delta \mathrm{Q}$. In the special case where $\mathrm{X}_k$ and $\mathrm{Y}_k$ are symmetric, both $\mathrm{Q}_{\epsilon+}$ and $\mathrm{Q}_{\epsilon-}$ degenerate to zero matrices. 
		This can be verified by checking that $\|\mathbf{Z}_{+}\| \le 1$ if $\Delta \mathrm{Q} \succeq \mathbf{0}$ and $\|\mathbf{Z}_{-}\| \le 1$ if $\Delta \mathrm{Q} \preceq \mathbf{0}$ under Assumption \ref{assump2}. 
		In the more general case where $\mathrm{X}_k$ and $\mathrm{Y}_k$ are not symmetric, one can determine $\Delta Q \succeq \mathrm{Q}_{\epsilon+}$ (or $\Delta Q \preceq \mathrm{Q}_{\epsilon-}$) by verifying the conditions in Lemma \ref{lemmazz}. Once satisfied, we obtain $(\mathrm{X}_k+\mathrm{Y}_k)(\mathrm{X}_k+\mathrm{Y}_k)^{T}-\mathrm{X}_k \mathrm{X}_k^{T} \succeq 0$ if $\Delta \mathrm{Q} \succeq \mathrm{Q}_{\epsilon+}$ and $(\mathrm{X}_k+\mathrm{Y}_k)(\mathrm{X}_k+\mathrm{Y}_k)^{T}-\mathrm{X}_k \mathrm{X}_k^{T} \preceq 0$ if $\Delta \mathrm{Q} \preceq \mathrm{Q}_{\epsilon-}$. It is worth noting that, unlike $\mathrm{Q}_{\epsilon+}$, $\mathrm{Q}_{\epsilon-}$ is not always attainable because of the restriction $\mathrm{Q}_k^{u}=\mathrm{Q}_k+ \Delta \mathrm{Q} \succeq 0$ in the Kalman filter.
	\end{remark}
	\begin{remark}
		One illustrative example of symmetric $\mathrm{X}_k$ and $\mathrm{Y}_k$ arises in a two-dimensional tracking problem where $\mathrm{H}_k=\mathrm{I}$, as shown in example 1 of \citep{b25}. When the measurement matrix is a scalar matrix, i.e., $\mathrm{R}_k=c^{-1}\mathrm{I}$, it follows that $\mathrm{H}_k^{T}\mathrm{R}_k^{-1}\mathrm{H}_k=c\mathrm{I}$. In this case, both $\mathrm{X}_k$ and $\mathrm{Y}_k$ are symmetric because the identity matrix commutes with any symmetric matrix. In more general scenarios, where $\mathrm{H}_k^{T}\mathrm{R}_k^{-1}\mathrm{H}_k \neq c\mathrm{I}$, the matrices $\mathrm{X}_k$ and $\mathrm{Y}_k$ are generally non-symmetric.
	\end{remark}
	
	\begin{theorem}
		If KF is stable and both $\mathrm{Q}_{\epsilon+}$ and $\mathrm{Q}_{\epsilon-}$ exist, it follows that $C_{\gamma,k}^{o} \ge C_{\gamma,k}^{u}$ whenever $\Delta \mathrm{Q} \succeq \mathrm{Q}_{\epsilon+}$, and $C_{\gamma,k}^{o} \le C_{\gamma,k}^{u}$ whenever $\Delta \mathrm{Q} \preceq \mathrm{Q}_{\epsilon-}$.
		\label{theorem1}
	\end{theorem}
	The proof is available in Appendix \ref{appendix5}. 
	\begin{remark}
		Theorem \ref{theorem1} indicates that the convergence of the improper initial guess would be accelerated when applying $\Delta Q \succeq  \mathrm{Q}_{\epsilon+}$ and decelerated when applying $\Delta Q \preceq \mathrm{Q}_{\epsilon-}$. \textcolor{black}{Furthermore, if Assumption \ref{assump2} holds, Theorem \ref{theorem1}  becomes $C_{\gamma,k}^{o} \ge C_{\gamma,k}^{u}$ whenever  $\Delta \mathrm{Q} \succeq \textbf{0}$ and $C_{\gamma,k}^{o} \le C_{\gamma,k}^{u}$ whenever  $\Delta \mathrm{Q} \preceq \textbf{0}$. }
		\label{remark4}
	\end{remark}
	
	\begin{prop}[Infinite Convergence Rate]
		In the case that the eigenvalues of $\Delta \mathrm{Q}$ tend to infinity, i.e., $\Delta \mathrm{Q} \to \infty$ and $\mathrm{H}_k^{T} \mathrm{R}_k^{-1} \mathrm{H}_k$ is a positive definite (PD) matrix, \textcolor{black}{$\mathrm{\hat{x}}_k^{b}$  converges to zero with an infinite convergence rate, i.e., $\mathrm{\hat{x}}_1^{b}=0$ for any arbitrary $\bar{x}_0^{\mathrm{b}}$.}
		\label{prop3}
	\end{prop}
	The proof is available in Appendix \ref{appendix6}.
	\subsubsection{Effects of $\Delta \mathrm{Q}$ on the Error Covariance}
	We investigate effects of $\Delta \mathrm{Q}$ on the error covariance in KF by analyzing the relations of the following three performance indices: ideal error covariance $\mathrm{P}_{k|k}$, filter calculated error covariance $\mathrm{P}_{k|k}^{f}$, and true error covariance $\mathrm{P}_{k|k}^{t}$. Note that $\mathrm{P}_{k|k}^{f}$ does not provide a true measure of estimation accuracy when $\Delta \mathrm{Q} \neq \textbf{0}$. 
	
	\begin{assumption}
		$\mathrm{P}_{k-1|k-1}=\mathrm{P}_{k-1|k-1}^{f}=\mathrm{P}_{k-1|k-1}^{t}$ at time step $k-1$.
		\label{assumption2}
	\end{assumption}
	The above assumption follows \citet{c13}. Accordingly, $\mathrm{P}_{k|k}$ can be obtained by the ideal KF as shown in \eqref{pk}. Under Assumption \ref{assumption2}, $\mathrm{P}_{k|k}^{f}$ can be calculated by 
	\begin{equation}
		\begin{aligned}
			\mathrm{\hat{x}}_{k|k}^{f} &= \mathrm{\hat{x}}_{k|k-1}^{f}+\mathrm{K}_k^{f}(\mathrm{y}_k-\mathrm{H}_k \mathrm{\hat{x}}_{k|k-1}^{f}) \\
			\mathrm{P}_{k|k}^{f}&=(\mathrm{I}-\mathrm{K}_k \mathrm{H}_k ) \mathrm{P}_{k|k-1}^{f} (\mathrm{I}-\mathrm{K}_k \mathrm{H}_k )^{T} + \mathrm{K}_k^{f} \mathrm{R}_{k} (\mathrm{K}_k^{f})^{T}\\
			\mathrm{\hat{x}}_{k|k-1}^{f} &= \Phi_k \mathrm{\hat{x}}_{k-1|k-1}\\
			\mathrm{P}_{k|k-1}^{f} &= \Phi_k \mathrm{P}_{k-1|k-1} \Phi_k^{T}  +  \mathrm{Q}_k^{u}\\
			\mathrm{K}_k^{f} &= \mathrm{P}_{k|k-1}^{f} \mathrm{H}_k^{T} (\mathrm{H}_k \mathrm{P}_{k|k-1}^{f} \mathrm{H}_k^{T}+\mathrm{R}_{k})^{-1}.
			\label{pkf}
		\end{aligned}
	\end{equation}
	The information form of $\mathrm{P}_{k|k}^{f}$ has
	\begin{equation}
		(\mathrm{P}_{k|k}^{f})^{-1}=(\mathrm{P}_{k|k-1}^{f})^{-1} + \mathrm{H}_k^{T} \mathrm{R}_k^{-1} \mathrm{H}_k.
		\label{infpf}
	\end{equation}
	Note that $\mathrm{P}_{k|k}^{f}$ does not provide a true measure of the estimation accuracy at time step $k$. Since
	\begin{equation}\nonumber
		\begin{aligned}
			\mathrm{\tilde{x}}_{k|k}^{f} &= \mathrm{x}_k - \mathrm{\hat{x}}_{k|k}^{f} \\
			&=(\mathrm{I}-\mathrm{K}_k^{f} \mathrm{H}_k )\Phi_k \mathrm{\tilde{x}}_{k-1|k-1}+(\mathrm{I}-\mathrm{K}_k^{f} \mathrm{H}_k )\mathrm{w}_k- \mathrm{K}_k^{f} \mathrm{v}_k,
		\end{aligned}
	\end{equation}
	it follows that the true error covariance $\mathrm{P}_{k|k}^{t}=\mathrm{cov}(\mathrm{\tilde{x}}_{k|k}^{f})$ has
	\begin{equation}
		\mathrm{P}_{k|k}^{t}=(\mathrm{I}-\mathrm{K}_k^{f} \mathrm{H}_k ) \mathrm{P}_{k|k-1} (\mathrm{I}-\mathrm{K}_k^{f} \mathrm{H}_k )^{T} + \mathrm{K}_k^{f} \mathrm{R}_{k} \mathrm{K}_k^{T}.
		\label{pkt}
	\end{equation}
	Then, we have the following lemma and theorem.
	\begin{lemma} \citep{c13}
		Under Assumption \ref{assumption2}, if $\Delta \mathrm{Q} \succeq \mathrm{0}, \mathrm{P}_{k|k}^{f} \succeq \mathrm{P}_{k|k}^{t} \succeq \mathrm{P}_{k|k}$. Otherwise, if $- \mathrm{Q}_k\preceq \Delta \mathrm{Q} \preceq \mathrm{0}, \mathrm{P}_{k|k}^{t} \succeq \mathrm{P}_{k|k} \succeq \mathrm{P}_{k|k}^{f}$.
		\label{lemma6}
	\end{lemma}
	\begin{remark}
		Lemma \ref{lemma6} reveals that the true error covariance is always \textcolor{black}{larger} than the ideal error covariance $\mathrm{P}_{k|k}$ as long as $\Delta \mathrm{Q} \neq \textbf{0}$.
	\end{remark}
	\begin{theorem}
		Under Assumption \ref{assumption2}, if $\mathbf{0} \preceq \Delta \mathrm{Q}_1 \preceq \Delta \mathrm{Q}_2$, one has $\mathrm{P}_{k|k} \preceq\mathrm{P}_{k|k}^{f1} \preceq \mathrm{P}_{k|k}^{f2}$ and $\mathrm{P}_{k|k} \preceq \mathrm{P}_{k|k}^{t1} \preceq \mathrm{P}_{k|k}^{t2}$ where superscripts $1$ and $2$ denote corresponding results of applying $\Delta \mathrm{Q}_1$ and $\Delta \mathrm{Q}_2$. On the contrary, if~$\mathbf{0} \succeq \Delta \mathrm{Q}_1 \succeq \Delta \mathrm{Q}_2 \succeq - \mathrm{Q}_k$, one has $\mathrm{P}_{k|k}\succeq \mathrm{P}_{k|k}^{f1} \succeq \mathrm{P}_{k|k}^{f2}$ and $\mathrm{P}_{k|k} \preceq \mathrm{P}_{k|k}^{t1} \preceq \mathrm{P}_{k|k}^{t2}$.
		\label{theorem2}
	\end{theorem}
	The proof is available in Appendix \ref{appendix7}. 
	\begin{remark}
		According to Theorems \ref{theorem1} and \ref{theorem2}, one observes that the two targets, the bias convergence rate and the practical error covariance, are conflicting. A bigger process covariance would lead to a faster bias convergence speed, at the cost of enlarged error covariance. 
	\end{remark}
	
	\subsection{Bias-Variance Trade-off in KF-DOB}
	As indicated in Remark \ref{remark1}, the incorrect initial guess in KF can be generalized to the intermittent state jump in KF-DOB under step-like disturbance  (note that the disturbance itself is a state in KF-DOB). Before proceeding, we give the following assumptions.
	\begin{assumption}
		The disturbance in \eqref{sys} is the summation of a step signal plus a white noise term, i.e.,
		\begin{equation}
			d_k=\mathbf{1} \sum_{i=0}^n \alpha_i \chi_{A_i}(k) + w_{d,k}
			\label{stepd}
		\end{equation}
		where $\mathbf{1}$ denotes a vector with elements 1 of proper dimension, $w_{d,k} \sim \mathcal{N}(0,Q_d)$, $\alpha_i$ are real numbers, $A_i$ are intervals with $A_i \cap A_j=\emptyset$ and $\cup_{i=0}^n A_i=\mathbb{R}$, $\chi_{A_i}(k)$ is a indicator function of $k$:
		$$
		\chi_{A_i}(k)= \begin{cases}
			1,  \text { if } k \in A_i \\ 
			0,  \text { if } k \notin A_i \end{cases}.
		$$
		\label{assump3} 
	\end{assumption}
	\begin{assumption}
		The KF-DOB converges to its steady state before the next disturbance jump comes. 
		\label{kfdobcov}	   
	\end{assumption} 
	\begin{remark}
		A typical example of step-like disturbance is the DC motor with unknown loads ~\citep{aravkin2017generalized}. Moreover, it is common to apply step signals in convergence analysis in the adaptive filtering and control communities~\citep{haykin2002adaptive}. Under Assumptions \ref{assump3} and \ref{kfdobcov}, the problem KF-DOB with step-like disturbance is identical to the problem of KF with incorrect initial mean, except for the state jump happening intermittently in KF-DOB but only happening at $k=0$ for KF with improper initial mean.  
	\end{remark}
	
	We assume that there is no cross-correlation between the state and disturbance, i.e., $\mathrm{Q}_k=\operatorname{diag}(\mathrm{Q}_{d},\mathrm{Q}_{x})$ for \eqref{sys}. Since the state jump only occurs on the disturbance part, we use the following process covariance 
	\begin{equation}
		\mathrm{Q}_k^{u}=\begin{bmatrix}
			\mathrm{Q}_{d}+	\Delta \mathrm{Q}_{d}&0\\
			0&\mathrm{Q}_{x}
		\end{bmatrix}.
		\label{covu}
	\end{equation} 
	Subsequently, we investigate the influence of $\Delta \mathrm{Q}_{d}$ on the bias-variance effects of KF-DOB. To simplify the analysis, we assume that the disturbance switched from $\mathbf{1}\alpha_1$ to $\mathbf{1}\alpha_{2}$ at time step $j+1$, i.e., $d_{j}=\mathbf{1}\alpha_1+w_{d_{j}}$ while $d_{j+1}=\mathbf{1}\alpha_{2}+w_{d_{j+1}}$. Correspondingly, at time step $j+1$, the ``initialization error" becomes
	\begin{equation}
		\begin{aligned}
			\bar{\mathrm{x}}_j^{\mathrm{b,dob}}&=[(d_{j}^\mathrm{b,dob})^{T},(x_{j}^\mathrm{b,dob})^{T}]\\
			&=[(\alpha_{2}-\alpha_{1})\mathbf{1}^{T},\mathbf{0}^{T}]^{T}.
			\label{inierror}
		\end{aligned}
	\end{equation}
	This error converges to zero as $j \to \infty$ if KF-DOB is stable. Consequently, we denote the true steady-state error covariance as $\mathrm{P}_{\infty}^{t,dob}$. 
	\begin{theorem}
		There exists an intrinsic bias-variance trade-off in KF-DOB associated with $\Delta \mathrm{Q}_d$. Under Assumption \ref{assump2}, choosing $\Delta \mathrm{Q}_d = 0$ minimizes the steady-state covariance $\mathrm{P}_{\infty}^{t, dob}$, aligning it with the ideal error covariance. However, this choice leads to the slowest convergence rate of the bias $\bar{x}_j^{\mathrm{b,dob}}$ to zero. Conversely, as $\Delta \mathrm{Q_d} \to \infty$, the convergence rate of the bias $\bar{x}_j^{\mathrm{b,dob}}$ is maximized, but this comes at the expense of increasing the steady-state covariance $\mathrm{P}_{\infty}^{t,dob}$.
		\label{theorem3}
	\end{theorem}
	The proof of this theorem is available in Appendix \ref{proofth3}. 
	\begin{theorem}
		Consider a more general process covariance mismatch $\Delta \mathrm{Q}=\begin{bmatrix}
			\Delta \mathrm{Q}_{d} & 0\\
			0 & \Delta \mathrm{Q}_x 
		\end{bmatrix}$. In this scenario, KF-DOB is an unbiased minimum variance estimator if and only if $\Delta \mathrm{Q}_d \to \infty$ and $\mathrm{Q}_{x}=0$. 
		\label{theorem4}
	\end{theorem}
	The proof of this theorem is available in Appendix \ref{proofth4}. 
	\section{Remedies}
	In this section, we build a connection between SISE and conventional KF-DOB. We prove that these estimators are identical when selecting infinite disturbance noise covariance. Moreover, we demonstrate that KF-DOB is insufficient for complex disturbance scenarios. To handle this issue, we provide two remedies, i.e., MKCKF-DOB and IMMKF-DOB.    
	\subsection{Native Kalman Filter-based Disturbance Observer}	
	\textcolor{black}{ We consider the following NKF-DOB as follows:
		\begin{equation}
			\begin{aligned}
				\begin{bmatrix}
					d_{k} \\
					x_{k}
				\end{bmatrix} &= \begin{bmatrix}
					0   &0\\
					G_k &F_k 
				\end{bmatrix}\begin{bmatrix}
					d_{k-1} \\
					x_{k-1}
				\end{bmatrix}+ \begin{bmatrix}
					w_{d,k-1} \\
					w_{x,k}
				\end{bmatrix}\\
				{y}_k&=\begin{bmatrix}
					{0} & H_k
				\end{bmatrix} \begin{bmatrix}
					d_{k} \\
					x_{k}
				\end{bmatrix}+{v}_{k}.
			\end{aligned}
			\label{nkf-dob}
	\end{equation}}
	where $w_{d,k-1} \sim \mathcal{N}(0,D_k)$. The cross covariance has $E[(x_k-E(x_k))(d_{k-1}-E(d_{k-1}))^{T}]=E[(G_k d_{k-1}+F_k x_{k-1} + w_{x,k} -\bar{x}_k)(w_{d,k-1}-\bar{d}_{k-1})]=G_kD_k$. Analogically, one obtains $$
	E(\begin{bmatrix}
		w_{d,k-1} \\
		w_{x,k}
	\end{bmatrix}\begin{bmatrix}
		w_{d,k-1} \\
		w_{x,k}
	\end{bmatrix}^{T})=\begin{bmatrix}
		D_k & D_k G_{k}^{T}\\
		G_{k}D_k & Q_k
	\end{bmatrix}.$$
	\begin{assumption} 
		The initial state $x_0$ is independent of $w_{x,k}$ and $w_{d,k-1}$ with a known mean $x_{0|0}$ and covariance matrix $P_{0|0}^{x}$. Moreover, rank~$H_k G_k=\text{rank~} G_k =p$ where $p$ is the dimension of $d_k$.
		\label{assumption5}
	\end{assumption}
	Under Assumption \ref{assumption5} and applying the Gaussian conditional density formula, one arrives
	\begin{equation}
		\begin{aligned}
			P_{k|k-1}&=G_k D_k G_k^{T} + F_k P_{k-1|k-1} F_{k}^{T}+Q_k\\
			K_k&=P_{k|k-1}H_k^{T} (H_k P_{k|k-1} H_k^{T}+R_k)^{-1}\\
			M_{k}&=D_k G_{k}^{T}H_k^{T}(H_k P_{k|k-1} H_k^{T}+R_k)^{-1}\\
			P_{k|k}&=\mathrm{cov}(x_{k}|\{y_k\})=(I-K_{k}H_k)P_{k|k-1}\\
			P_{k-1|k}^{d}&=\mathrm{cov}(d_{k-1}|\{y_k\})=(I-M_k H_k G_k) D_k\\
			x_{k|k}&=F_k x_{k-1|k-1} + K_k(y_k-H_k F_k x_{k-1|k-1})\\
			d_{k-1|k} &= M_{k}(y_k-H_k F_k x_{k-1|k-1})\\
		\end{aligned}
		\label{nkfdob}
	\end{equation}
	where $d_{k-1|k}$ is the estimate of $d_{k-1}$ at time step $k$.
	\begin{lemma}\citep{b20}
		The NKFDOB in \eqref{nkfdob} is identical to SISE as shown in \eqref{sise1}-\eqref{sise3} when $D_{k} \to \infty$.
		\label{lemma33}
	\end{lemma}
	\begin{lemma}(Identity 9 of ~\citet{b20})
		$M_k \to M_k^{*}$ and $K_k \to G_k M_k^{*} + K_k^{*}(I-H_k G_k M_k^{*}) $ as $D_k \to \infty$. 
		\label{lemma9}
	\end{lemma}
	\subsection{Kalman filter-based Disturbance Observer}
	We consider the KFDOB as shown in \eqref{sys}. The disturbance noise is assumed to follow $w_{d,k-1} \sim \mathcal{N}(0,D_k)$. Note that the only difference between \eqref{nkf-dob} and \eqref{sys} is that a nominal model $d_k=d_{k-1}$ is used in \eqref{sys} but ignored in \eqref{nkf-dob}. In KF, denoting 
	$${P}_{k-1|k-1}=\begin{bmatrix}
		P^{dd} & P^{dx} \\
		P^{xd} & P^{xx} 
	\end{bmatrix}, \tilde{Q}_k =\begin{bmatrix}
		D_k & 0\\
		0 & Q_k
	\end{bmatrix},$$ one has
	\begin{equation}
		\scriptsize
		\begin{aligned}
			P_{k|k-1}^{dd}=&P^{dd}+D_k,~P_{k|k-1}^{dx}= P^{dd}G_k^{T} + P^{dx} F_k^{T} \\
			P_{k|k-1}^{xx}=& G_k P^{dd} G_k^{T} + F_k P^{xd} G_k^{T} + G_k P^{dx} F_k^{T}+ F_k P^{xx} F_k ^{T} + Q_k\\
			K_k^{x}=&P_{k|k-1}^{xx}H_k^{T} (H_k P_{k|k-1}^{xx} H_k^{T}+R_k)^{-1}\\
			M_{k}^{d}=&P_{k|k-1}^{dx}H_k^{T}(H_k P_{k|k-1}^{xx} H_k^{T}+R_k)^{-1}\\
			P_{k|k}^{xx}=&\mathrm{cov}(x_{k}|\{y_k\})=(I-K_{k}H_k)P_{k|k-1}^{xx}\\
			P_{k|k}^{dd}=&\mathrm{cov}(d_{k-1}|\{y_k\})=P_{k|k-1}^{dd}- P_{k|k-1}^{dx}H_k^{T}\\
			&\times(H_k P_{k|k-1}^{xx} H_k^{T}+R_k)^{-1} H_k  P_{k|k-1}^{xd}\\
			P_{k|k}^{dx}=&P_{k|k-1}^{dx}-P_{k|k-1}^{dx}H^{T}(H_k P_{k|k-1}^{xx} H_k^{T}+R)^{-1} H_k P_{k|k-1}^{xx}\\
			x_{k|k}=&F_k x_{k-1|k-1} + G_k d_{k-1|k-1}+ K_k^{x}(y_k\\
			&-H_k F_k x_{k-1|k-1}- H_k G_k d_{k-1|k-1})\\
			d_{k|k} =& d_{k-1|k-1}+  M_{k}(y_k-H_k F_k x_{k-1|k-1}-H_k G_k d_{k-1|k-1})\\
		\end{aligned}
		\label{kfdob}
	\end{equation}
	\begin{theorem}
		The KFDOB in \eqref{kfdob} is identical to NKFDOB in \eqref{nkfdob} as $D_k \to \infty$. 
		\label{theorem5}
	\end{theorem}
	The proof of this theorem is available in Appendix \ref{pftheorem5}.
	\begin{corollary}
		According to Lemma \ref{lemma33} and Theorem \ref{theorem5}, KFDOB is identical to the SISE estimator as $D_k \to \infty$. 
		\label{corollary3}
	\end{corollary}
	\begin{remark}
		Corollary \ref{corollary3} coincides with Theorem \ref{theorem4} which states that KF-DOB becomes an unbiased minimum variance estimator as $D_k \to \infty$ under arbitrary disturbance models.
	\end{remark}
	According to Corollary \ref{corollary3}, as $D_k \to \infty$, NKF-DOB, KF-DOB, and SISE coincide. As pointed out by ~\citep{b19}, SISE is an unbiased minimum variance estimator under an arbitrary disturbance model. Meanwhile, KF-DOB is an optimal estimator when the disturbance model is exactly known. In many practical applications, the disturbance model is complex and cannot be modeled accurately. In this case, based on Theorem \ref{theorem3}, there is an intrinsic bias-variance trade-off in KF-DOB. To alleviate this issue, we give two remedies: MKCKF-DOB and IMMKF-DOB. The former alleviates this trade-off through a robust loss, i.e., MKCL, while the latter uses a switching disturbance covariance to balance the disturbance tracking speed and tracking uncertainty. 
	\subsection{Multi-Kernel Correntropy Kalman Filter-Based Disturbance Observer}
	One can rewrite \eqref{sys} as
	\begin{equation}
		\mathrm{T}_k=\mathrm{W}_k \mathrm{x}_k + \mathrm{B}_{k}^{-1}{\nu}_k
	\end{equation}
	where 
	$$\scriptsize
	\mathrm{T}_k=\mathrm{B}_{k}^{-1}\begin{bmatrix}
		\mathrm{\hat{x}}_{k|k-1}\\
		\mathrm{y}_k
	\end{bmatrix},~
	\mathrm{W}_k=\mathrm{B}_{k}^{-1}\begin{bmatrix}
		\mathrm{I}\\
		\mathrm{H}_k
		\label{TW}
	\end{bmatrix},~{\nu}_k=\begin{bmatrix}
		\mathrm{\hat{x}}_{k|k-1}-\mathrm{x}_k\\
		\mathrm{v}_k
	\end{bmatrix}
	$$
	and $\mathrm{B}_{k}$ is obtained by Cholesky decomposition with 
	$$
	\begin{aligned}
		E({\nu}_k {\nu}_k^{T})&=\begin{bmatrix}
			\mathrm{P}_{k|k-1}&0\\
			0&\mathrm{R_k}
		\end{bmatrix}=\begin{bmatrix}
			\mathrm{B}_{p}\mathrm{B}_{p}^{T}&0\\
			0&\mathrm{B}_{r}\mathrm{B}_{r}^{T}
		\end{bmatrix}=\mathrm{B}_{k}\mathrm{B}_{k}^{T}.
	\end{aligned}
	\label{bpbr}
	$$
	MKCKF is obtained by minimizing the following MKCL~\citep{b27}:
	\begin{equation}
		\min_{\mathrm{x}_k} J_{MKCKF} =  \sum_{i=1}^{l} \sigma_{i}^{2}\big(1-G_{\sigma_{i}}(\mathrm{e}_{i,k})\big)
		\label{mkckf}
	\end{equation}
	where $l$ denotes the dimensions of $\mathrm{T}_k$, $\mathrm{e}_{i,k}$ denotes the $i$-th element of $\mathrm{e}_k$  with $\mathrm{e}_k=[\mathrm{e}_{p,k}^{T},\mathrm{e}_{r,k}^{T}]^{T}=\mathrm{T}_k-\mathrm{W}_k \mathrm{x}_k$, and $\sigma_i$ are kernel bandwidths.
	
	As a comparison, the KF can be derived by the following objective function~\citep{c12}:
	\begin{equation}
		\min_{\mathrm{x}_k} J_{KF}= \frac{1}{2}\|\mathrm{y}_k-\mathrm{H}_k \mathrm{x}_k\|_{{\mathrm{R}}^{-1}_k}^{2}+\frac{1}{2}\|\mathrm{x}_k-\mathrm{\Phi}_{k}\mathrm{x}_{k-1}\|_{\mathrm{P}_{k|k-1}^{-1}}^{2}.
		\label{kf}
	\end{equation}
	The loss function \eqref{kf} is actually identical to \eqref{mkckf} when $\sigma_i \to \infty, \quad \forall i$, establishing a connection between MKCKF and KF \citep{b27}. In a disturbance observer, we denote the process kernel bandwidth vector as $\boldsymbol{\sigma}_p=[\boldsymbol{\sigma}_d^{T},\boldsymbol{\sigma}_x^{T}]^{T}$, where $\boldsymbol{\sigma}_d$ and $\boldsymbol{\sigma}_x$ are the bandwidth vectors corresponding to the disturbance and the state, respectively. Furthermore, we denote $\boldsymbol{\sigma}_r$ as the kernel bandwidth vector associated with the measurement. Since only the nominal disturbance model $d_k=d_{k-1}$ may deviate from the true disturbance model, we set $\boldsymbol{\sigma}_x \to \infty$ and $\boldsymbol{\sigma}_r \to \infty$, and choose $\boldsymbol{\sigma}_d$ as a positive, tunable vector reflecting our confidence in the nominal disturbance model. Following a derivation similar to that presented in \citep{b27}, we summarize the MKCKF-DOB in Algorithm \ref{mkckf-dob}.
	\begin{algorithm}[t]
		\setstretch{0.9} 
		\caption{\textcolor{black}{MKCKF-DOB}}
		\label{mkckf-dob}
		\begin{algorithmic}[1]
			\State {\textbf{Step 1: Initialization}}\\
			Choose the bandwidth vector $\boldsymbol{\sigma}_d$ and a threshold $\varepsilon$.
			\State {\textbf{Step 2: State Prediction}}\\
			$\hat{\mathrm{x}}_{k|k-1}=\Phi_k \hat{\mathrm{x}}_{k-1|k-1} $\\
			$\mathrm{P}_{k|k-1}={\Phi_k} \mathrm{P}_{k-1|k-1 } {\Phi_k}^{T}+\mathrm{Q}_k$\\
			Obtain $\mathrm{B}_{p}$ and $\mathrm{B}_{r}$ by Cholesky decomposition.
			\State {\textbf{Step 3: State Update}}\\
			$\hat{\mathrm{x}}_{k|k,0}=\hat{\mathrm{x}}_{k|k-1}$
			\While{$\frac{\left\|\hat{\mathrm{x}}_{k|k,t}-\hat{\mathrm{x}}_{k|k,t-1}\right\|}{\left\|\hat{\mathrm{x}}_{k|k,t}\right\|}>\varepsilon$} \\
			$\hat{\mathrm{x}}_{k|k,t}=\hat{\mathrm{x}}_{k|k-1}+\tilde{\mathrm{K}}_{k,t}(\mathrm{y}_k-\mathrm{H}_k\hat{\mathrm{x}}_{k|k-1})$ \\
			$\tilde{\mathrm{K}}_{k,t}=\tilde{\mathrm{P}}_{ k|k-1}\mathrm{H}_k^{T}(\mathrm{H}_k\tilde{\mathrm{P}}_{ k|k-1}\mathrm{H}_k^{T}+\tilde{\mathrm{R}}_{k})^{-1}$\\
			$\tilde{\mathrm{P}}_{ k|k-1}=\mathrm{B}_{p}\tilde{\mathrm{M}}_{p}^{-1}{\mathrm{B}}_{p}^{T},~\tilde{\mathrm{R}}_{ k}=\mathrm{B}_{r}\tilde{\mathrm{M}}_{r}^{-1}\mathrm{B}_{r}^{T}$\\
			$ \tilde{\mathrm{M}}_{p}=\mathrm{diag}(\mathrm{G}_{\boldsymbol{\sigma}_p}(\mathrm{e}_{p,k})),~\tilde{\mathrm{M}}_{r}=\mathrm{diag}(\mathrm{G}_{\boldsymbol{\sigma}_r}(\mathrm{e}_{r,k}))$\\
			$\mathrm{e}_{p,k}=\mathrm{B}_{p}^{-1}\hat{\mathrm{x}}_{k|k-1}-\mathrm{B}_{p}^{-1}\hat{\mathrm{x}}_{k|k,t-1}$\\
			$\mathrm{e}_{r,k}=\mathrm{B}_{r}^{-1}\mathrm{y}_{k}-\mathrm{B}_{r}^{-1}\mathrm{H}\hat{\mathrm{x}}_{k|k,t-1}$\\
			$t=t+1$
			\EndWhile \\
			$\mathrm{P}_{k|k}=(\mathrm{I}-\tilde{\mathrm{K}}_k \mathrm{H}_k){\mathrm{P}}_{k|k-1}(\mathrm{I}-\tilde{\mathrm{K}}_k \mathrm{H}_k)^{T}+\tilde{\mathrm{K}}_k \mathrm{R}_k\tilde{\mathrm{K}}_k^{T}$		
		\end{algorithmic}
	\end{algorithm}
	\begin{remark}
		In MKCKF-DOB, by applying $\boldsymbol{\sigma}_x \to \infty$ and $\boldsymbol{\sigma}_r \to \infty$, one has $\tilde{\mathrm{R}}_{k}={\mathrm{R}}_{k}$ and only the submatrix of ${\mathrm{P}}_{ k|k-1}$ associated with the disturbance is inflated where the inflation level is determined by the process error $\mathrm{e}_{p,k}$. This mechanism can be understood as applying an ``adaptive'' disturbance process covariance matrix so that it matches the practical process error under certain information metrics. 
		\label{remark9}
	\end{remark}
	
	\textcolor{black}{The convergence analysis of Algorithm \ref{mkckf-dob} has been established by Theorems 8 and 9 in \citet{b27}, which states that all elements of $\boldsymbol{\sigma}_d$ should be larger than a certain constant so that the fixed-point iteration in Lines 9-17 of Algorithm \ref{mkckf-dob} converges. Note that MKCKF-DOB is identical to KF-DOB when applying sufficiently large $\boldsymbol{\sigma}_d$ and hence we advise tuning kernel bandwidths from a big value to a relatively small value gradually to avoid divergence at the beginning. It is also observed that the algorithm complexity of MKCKF-DOB is moderately higher than KF-DOB, and the complexities of the whole algorithm depend on the average iteration number $\bar{t}$. Fortunately, $\bar{t}$ generally lies between 2 and 5 in MKCKF-DOB in most situations. One can see the detailed time consumption of different estimators in the Simulation Section.}

	\subsection{Interacting Multiple Model Kalman Filter-based Disturbance Observer}
	As demonstrated in Theorem \ref{theorem3}, there is an intrinsic bias-variance trade-off in KF-DOB regarding the disturbance noise covariance selection. This dilemma can be mitigated by designing a switching disturbance process covariance mechanism so that the disturbance tracking is timely and smooth. To achieve this purpose, we design an IMMKF-DOB that has a similar structure to the conventional IMMKF summarized in Section \ref{immkf}, whereas a unified dynamic model accompanied with different process covariance is utilized in  IMMKF-DOB. Specifically, in IMMKF-DOB, model $j$ has 
	\begin{equation}
		\begin{aligned}
			\mathrm{x}_k &= \Phi_k \mathrm{x}_{k-1} + \mathrm{w}_{j,k} \\
			\mathrm{y}_k &= \mathrm{H}_k \mathrm{x}_k  + \mathrm{v}_k
		\end{aligned}
		\label{sysimm}
	\end{equation}
	where $\mathrm{w}_{j,k}\sim \mathcal{N}\Big(0,\begin{bmatrix}
		\mathrm{Q}_{dj,k}&0\\
		0&\mathrm{Q}_{x,k}
	\end{bmatrix}\Big)$ and $\mathrm{Q}_{dj,k}$ is the disturbance noise covariance for model $j$. By selecting multiple $\mathrm{Q}_{dj,k}$, IMMKF-DOB can achieve a timely and smoothing disturbance tracking through the fusion of multiple models (i.e., multiple disturbance covariances). \textcolor{black}{The detailed IMMKF-DOB algorithm is similar to IMMKF~\citep{c9} as shown in equations \eqref{imm1} to \eqref{imm5} and is summarized in Algorithm \ref{immkf-dob}.}
	\begin{algorithm}[t]
		\setstretch{1} 
		\caption{{IMMKF-DOB}}
		\label{immkf-dob}
		\begin{algorithmic}[1]
			\State \textbf{Step 1: Initialization} \\
			Initialize the transition probability matrix $\mathcal{P}$ and disturbance noise covariance $\mathrm{Q}_{dj,k}$ for model $j=1,2,\ldots,q$.
			\State {\textbf{Step 2: Input Interaction}}\\
			Obtain the initial state and covariance estimates of model $j$ through \eqref{imm1} and \eqref{imm2}.
			\State {\textbf{Step 3: Filtering}}\\
			Run the KF for model $j$ through \eqref{imm3} and \eqref{imm31}.
			\State {\textbf{Step 4: Output Interaction}}\\
			Update the model probability and calculate the posterior state and error covariance through \eqref{imm4} and \eqref{imm5}.
		\end{algorithmic}
	\end{algorithm}
	
	\textcolor{black}{
		The convergence analysis of IMMKF is established by Theorem 4 of \citet{hwang2016study}, which states that the IMMKF is globally exponentially stable if the system dynamics are uniformly controllable and uniformly observable under identical observation models. This result ensures the stability of our proposed IMMKF-DOB, as it is a special case of the IMMKF. To effectively estimate the potentially complex disturbances, we recommend selecting at least two distinct disturbance noise covariance matrices, namely $\mathrm{Q}_{d1,k}$ and $\mathrm{Q}_{d2,k}$. The former should be set as a small positive semi-definite (PSD) matrix to produce smooth disturbance estimates, while the latter should be assigned as a larger PSD matrix to account for abrupt changes. The computational complexity of IMMKF-DOB is $\mathcal{O}(qn^3+q^2n^2)$, where $q$ is the model number and $n$ is the state dimension~\citep{bar2004estimation}; see the detailed time consumptions in the Simulation Section.}
	\section{Simulations}
	\textcolor{black}{Three examples, the vehicle tracking, 1-DOF torsion system, and 1-DOF manipulator example, are considered. The corresponding codes are \href{https://github.com/lsl-zsj/Bias-Variance-Trade-off-in-Kalman-Filter-Based-Disturbance-Observers}{available online}\footnote[1]{https://github.com/lsl-zsj/Bias-Variance-Trade-off-in-Kalman-Filter-Based-Disturbance-Observers}.}
	\subsection{\textcolor{black}{Vehicle Tracking Example}}
	We consider the following vehicle tracking problem:
	\begin{equation}
		\begin{aligned}
			x_k &= F_k x_{k-1} + G_k d_{k-1} + w_k\\
			y_k &= H_k x_k +v_k\\
		\end{aligned}
	\end{equation}
	where $F_k= \begin{bmatrix}
		1&T\\
		0&1
	\end{bmatrix}$, $G_k = \begin{bmatrix}
		\frac{T^2}{2}\\
		T
	\end{bmatrix}$, $H_k=\begin{bmatrix}
		1&0\\
		0&1
	\end{bmatrix}$, $x_k=[p_k ,v_k]^{T}$ which contains the position and velocity, $T=0.1$s is the sampling time, and $d_k$ is step-like disturbance as shown in \eqref{stepd}.
	\subsubsection{Trade-off Property}
	\label{tradeoff}
\begin{figure}[htbp]
		\centering 	
		\subfigure[NKF-DOB]{
			\begin{minipage}[t]{0.45\linewidth}
				\centering
				\includegraphics[width=1.0\columnwidth]{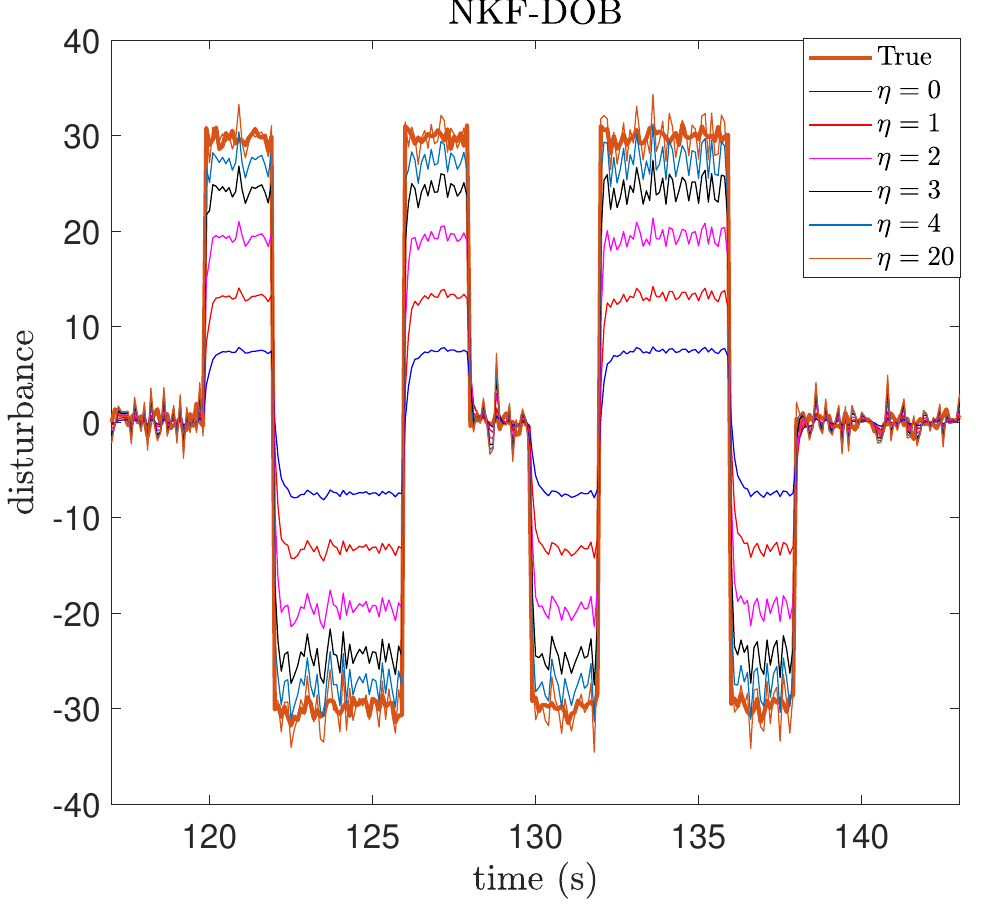}
				\label{nkf_d}
			\end{minipage}%
		}%
		\subfigure[KF-DOB]{
			\begin{minipage}[t]{0.45\linewidth}
				\centering
				\includegraphics[width=1.0\columnwidth]{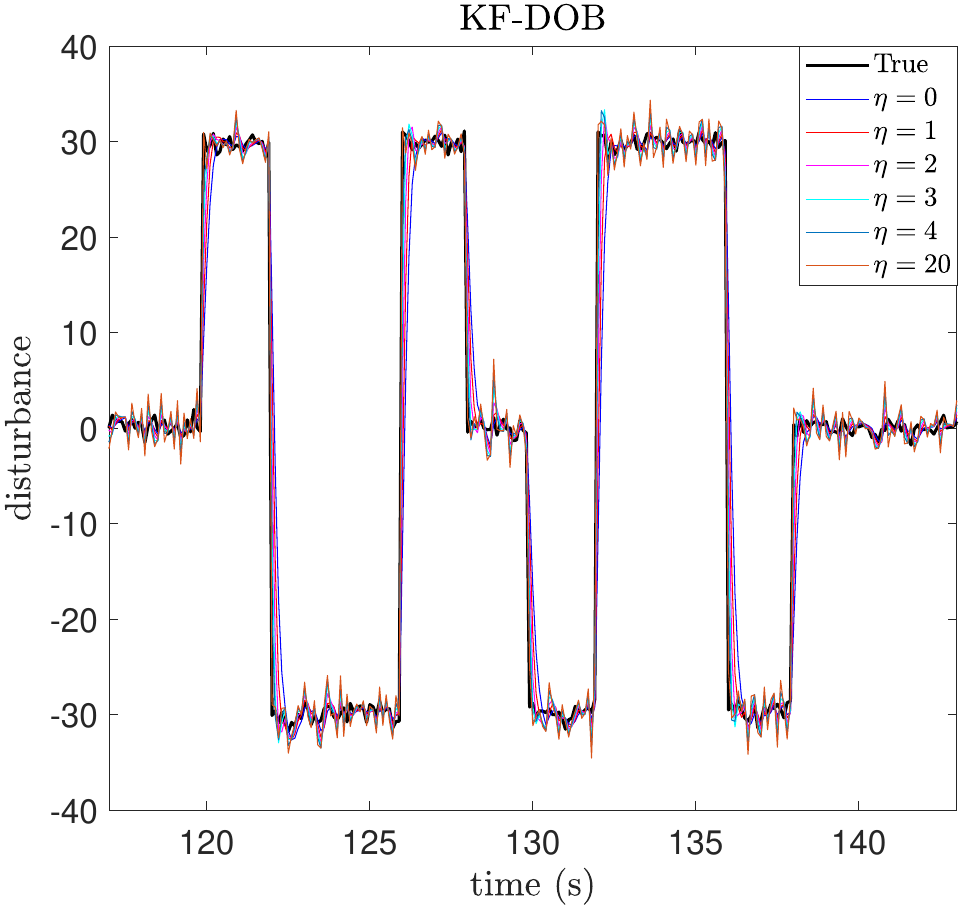}
				\label{akf_d}
			\end{minipage}%
		}%
		\caption{The disturbance estimation performances of NKF-DOB and KF-DOB with different $\eta$.}	
		\label{trade-off}
	\end{figure}

	We denote the nominal disturbance covariance as ${D}^{*}$. Then, we compare the performance of NKF-DOB and KF-DOB with $D=\exp(\eta) {D}^{*}$ where $\eta$ ranging from $0$ to $20$. The results are shown in Fig. \ref{trade-off}. One can observe that when $\exp(\eta)$ is finite and the disturbance is a step-like function, NKF-DOB and KF-DOB are biased but their estimate is relatively smooth. On the contrary, when $\exp(\eta)$ tends to infinity, the estimators are unbiased at the price of a non-smooth estimate. 
	
\begin{figure}[htbp]
		\centering 	
		\subfigure[]{
			\begin{minipage}[t]{0.45\linewidth}
				\centering
				\includegraphics[width=1.0\columnwidth]{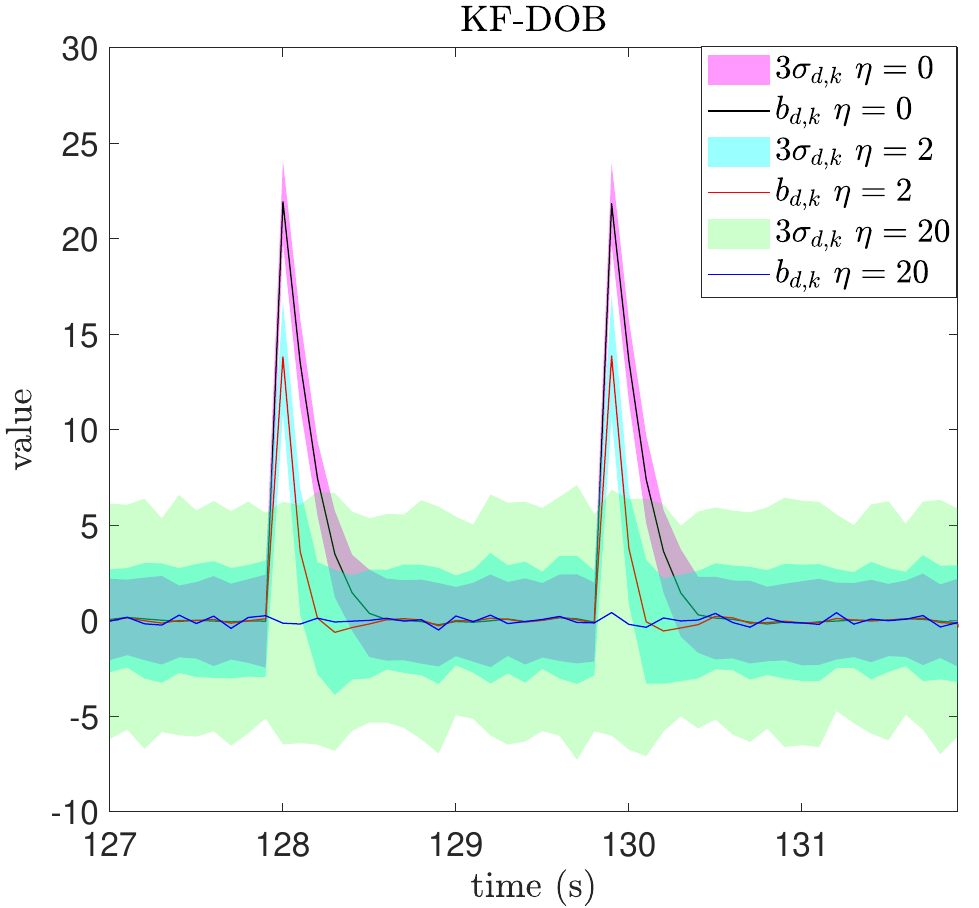}
				\label{tradeoff1}
			\end{minipage}%
		}%
		\subfigure[]{
			\begin{minipage}[t]{0.45\linewidth}
				\centering
				\includegraphics[width=1.0\columnwidth]{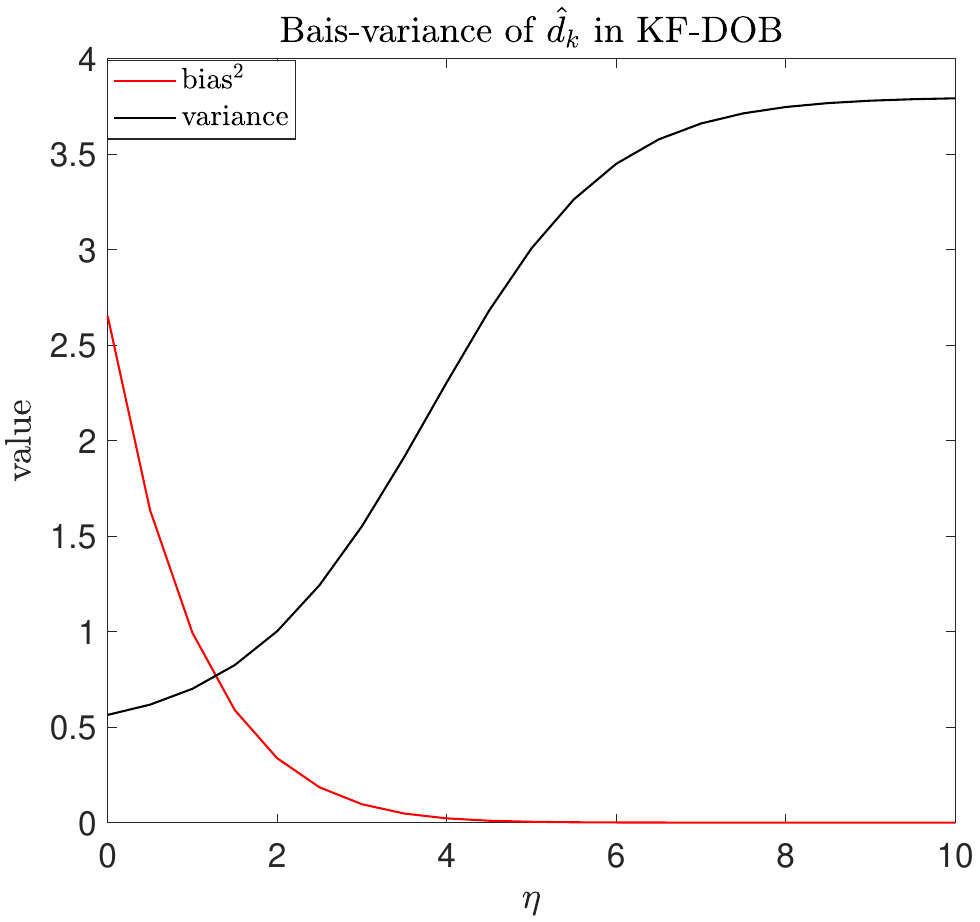}
				\label{bias-variance-kfdob}
			\end{minipage}%
		}%
		\caption{The visualization of bias-variance trade-off in KF-DOB with different $\eta$.}	
		\label{bias_variance_tradeoff}
	\end{figure}

	To highlight the bias-variance trade-off in KF-DOB, we conduct 100 Monte Carlo runs to compare the bias and standard deviation of the disturbance estimate at each time step with different $\eta$. It is worth mentioning that the average disturbance bias is obtained by $b_{d,k}= \frac{1}{K}\sum_{i=1}^{K} (d_k -\hat{d}_k)$ and the standard deviation is obtained by $\sigma_{d,k}=\sqrt{\frac{1}{K}{\sum_{i=1}^{K}(d_k -\hat{d}_k-b_{d,k})^{2}}}$, where $d_k$ is the ground truth disturbance, $\hat{d}_k$ is the estimated disturbance, and $K$ is the Monte Carlo counts. The results are shown in Fig. \ref{tradeoff1}. We observe that the $3\sigma_{d,k}$ region of $\eta=0$ is substantially tighter than that of  $\eta=20$, at the cost of increased bias when a step-like disturbance comes.  Meanwhile, when $\exp(\eta)$ is sufficiently large, the $3\sigma_{d,k}$ region becomes wider but the bias effects disappear gradually. We further summarize the average square of bias $\bar{b}_{d}^{2}=\frac{1}{m_2-m_1+1}\sum_{k=m_1}^{m_2} b_{d,k} $ and average variance $\bar{\sigma}^2=\frac{1}{m_2-m_1+1}\sum_{k=m_1}^{m_2} \sigma_{k}^2$ within the time interval $t=[126,132]$ (i.e., $m_1=1260$ and $m_2=1320$. This region contains both the constant disturbance region and disturbance jump region). The corresponding result is shown in Fig. \ref{bias-variance-kfdob}, where one can observe an obvious bias-variance trade-off with the increment of $\eta$ (i.e., the disturbance noise covariance). 
	
	\subsubsection{\textcolor{black}{Identity Property}} 
\begin{figure}[htbp]
		\centering
		\subfigure[]{
			\begin{minipage}[t]{0.45\linewidth}
				\centering
				\includegraphics[width=1.0\columnwidth]{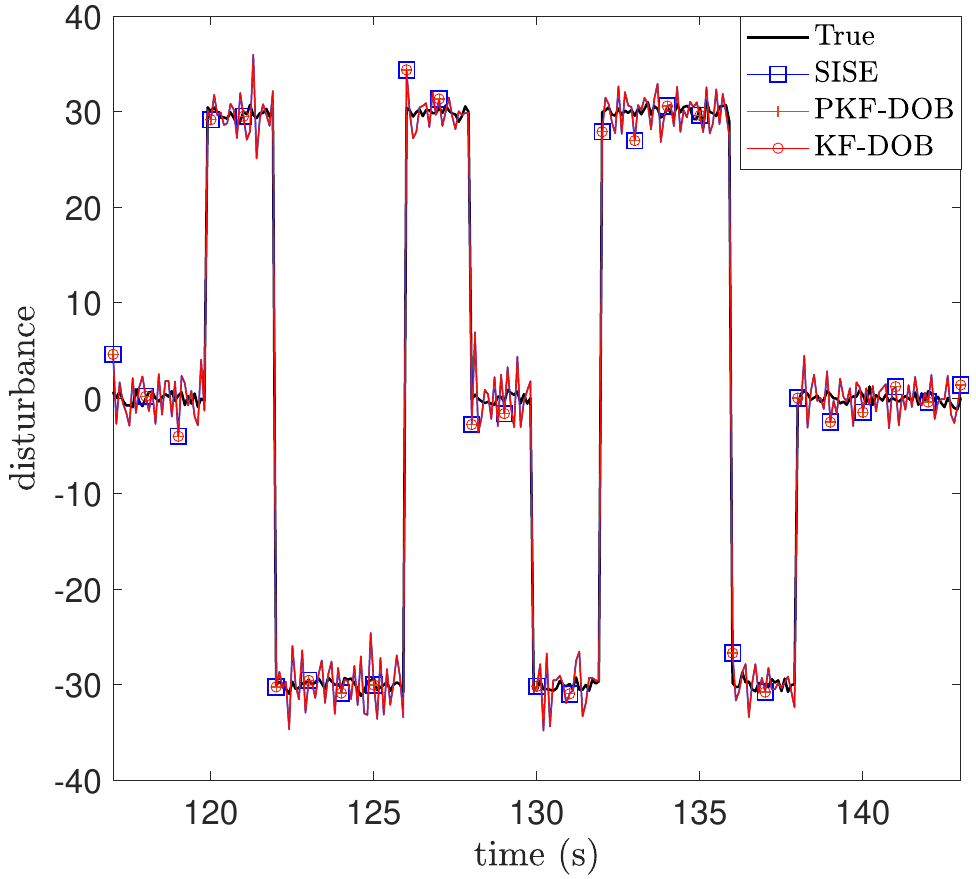}
				\label{dcompare}
			\end{minipage}%
		}%
		\subfigure[]{
			\begin{minipage}[t]{0.45\linewidth}
				\centering
				\includegraphics[width=1.0\columnwidth]{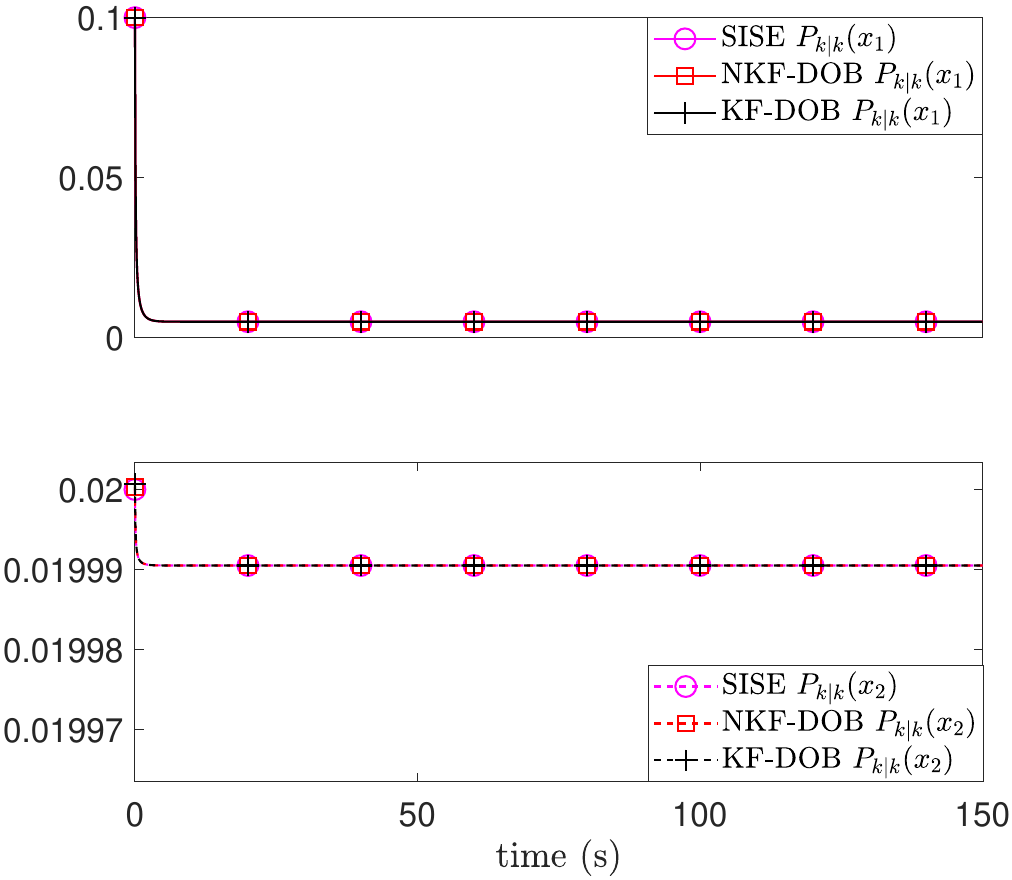}
				\label{covcompare}
			\end{minipage}%
		}%
		\caption{Identity property of SISE, NKF-DOB, and KF-DOB. (a) The disturbance estimate of different estimators. (d) Error covariance of different estimators.}
		\label{identity}	
	\end{figure}

	We apply a sufficiently large $D$ (e.g., $D=\exp(20)D^{*}$) for NKF-DOB and KF-DOB, and compare them with the SISE estimator. The results are shown in Fig. \ref{identity}. One can observe that these estimators are identical which coincides with Theorem \ref{theorem5} and Corollary \ref{corollary3}.
	\begin{table*}[]
		\centering
		\caption{Root-mean-square-error and Time Consumption of Different Estimators.}
		\scalebox{0.95}{
			\begin{tabular}{ccccc}
				\hline
				\hline
				Algorithm & $d-\hat{d}$ (mean$ \pm$ std ) & $x_1-\hat{x}_1$ (mean$ \pm$ std ) &  $x_2-\hat{x}_2$ (mean$ \pm$ std )& time cost (mean$ \pm$ std ) \\
				\hline
				KF-DOB $\eta=\exp(0)$& 2.2274 $\pm$0.0125 & 0.0674$\pm$ 0.004537 & 0.1743 $\pm$0.0024 & 0.0078 $\pm$0.0019 \\
				
				KF-DOB $\eta=\exp(1)$&1.7857$\pm$ 0.0137 & 0.0673$\pm$ 0.004548 & 0.1434$\pm$ 0.0022 & 0.0076 $\pm$0.0010 \\
				
				KF-DOB $\eta=\exp(2)$& 1.5008 $\pm$0.0158 & 0.0672 $\pm$0.004552 & 0.1327 $\pm$0.0019 & 0.0076$\pm$ 0.0017 \\
				
				KF-DOB $\eta=\exp(3)$&1.4497 $\pm$0.0190 & 0.0672 $\pm$0.004552 & 0.1325 $\pm$ 0.0017 & 0.0076 $\pm$0.0017 
				\\
				KF-DOB $\eta=\exp(20)$&2.0025 $\pm$0.0311 & 0.0672 $\pm$0.0045493 & 0.1417 $\pm$0.0016 & 0.0077 $\pm$0.0016\\
				
				MKCKF-DOB& 0.7500 $\pm$ 0.0105 & 0.0672 $\pm$ 0.004560 & 0.1102 $\pm$0.0016 & 0.0188 $\pm$0.0039 \\
				IMMKF-DOB& 0.9412 $\pm$0.0183 & 0.0671 $\pm$0.004556 & 0.1176 $\pm$0.0016 & 0.0296 $\pm$0.0027\\
				\hline
				\hline
		\end{tabular}}
		\label{simr}
	\end{table*}
	\subsubsection{\textcolor{black}{Two Remedies}}
\begin{figure}[htbp]
		\centering
		\includegraphics[width=0.5\columnwidth]{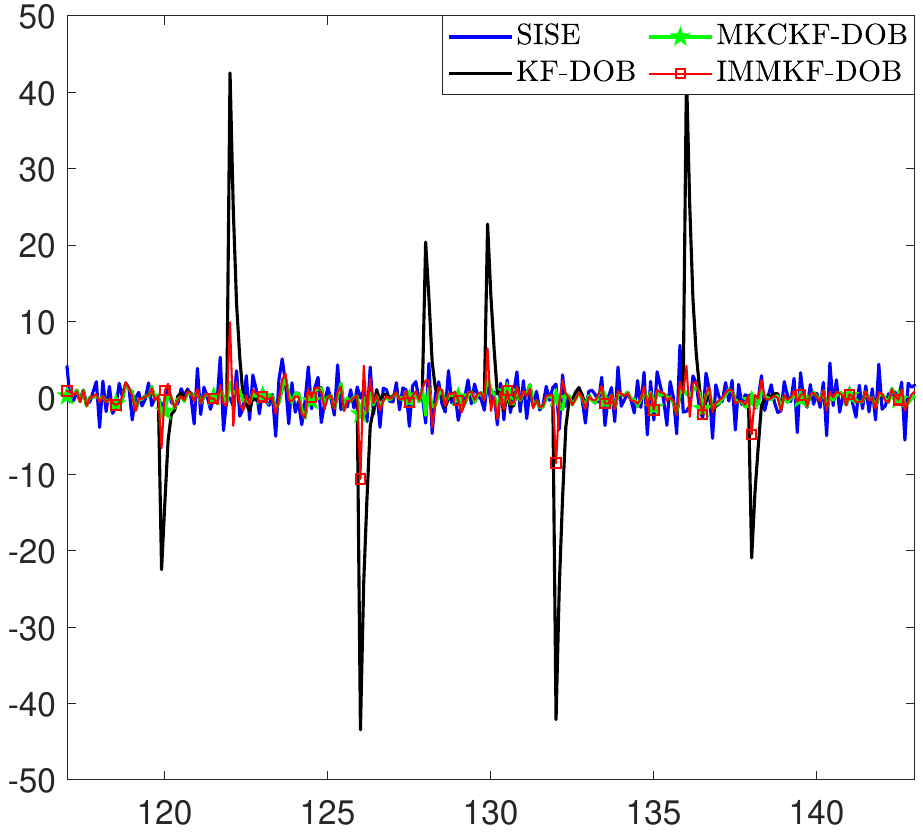}
		\caption{The disturbance estimate of SISE, KF-DOB, MKCKF-DOB, and IMMKF-DOB.}	
		\label{mkckfdob}
	\end{figure}

	In MKCKF-DOB, we use $\boldsymbol{\sigma}_p=[\varsigma_d,10^{8},10^{8}]^{T}$ with $\varsigma_d=3$ and $\boldsymbol{\sigma}_r=[10^{8},10^{8}]$ according to Remark \ref{remark9}. Moreover, we apply $D=D^{*}$ which is the same with the KF-DOB setting. As for IMMKF-DOB, we use $D_1=D^{*}$ for Model 1 and $D_2=\exp(5)D^{*}$ for Model 2. The Markov transition matrix has $\mathcal{P}=\begin{bmatrix}
		0.98&0.02\\
		0.5&0.5
	\end{bmatrix}$. The corresponding disturbance error performances of SISE, KF-DOB, MKCKF-DOB, and IMMKF-DOB are shown in Fig. \ref{mkckfdob}. One can observe that MKCKF-DOB and IMMKF-DOB have a smaller bias at the disturbance jump moments compared with the KF-DOB, but have a similar performance with the KF-DOB at the constant disturbance moments, which indicates that these two remedies have a better bias-variance trade-off compared with conventional KF-DOB.  
	\begin{figure}[htbp]
		\centering
		\subfigure[]{
			\begin{minipage}[t]{0.45\linewidth}
				\centering
				\includegraphics[width=1.0\columnwidth]{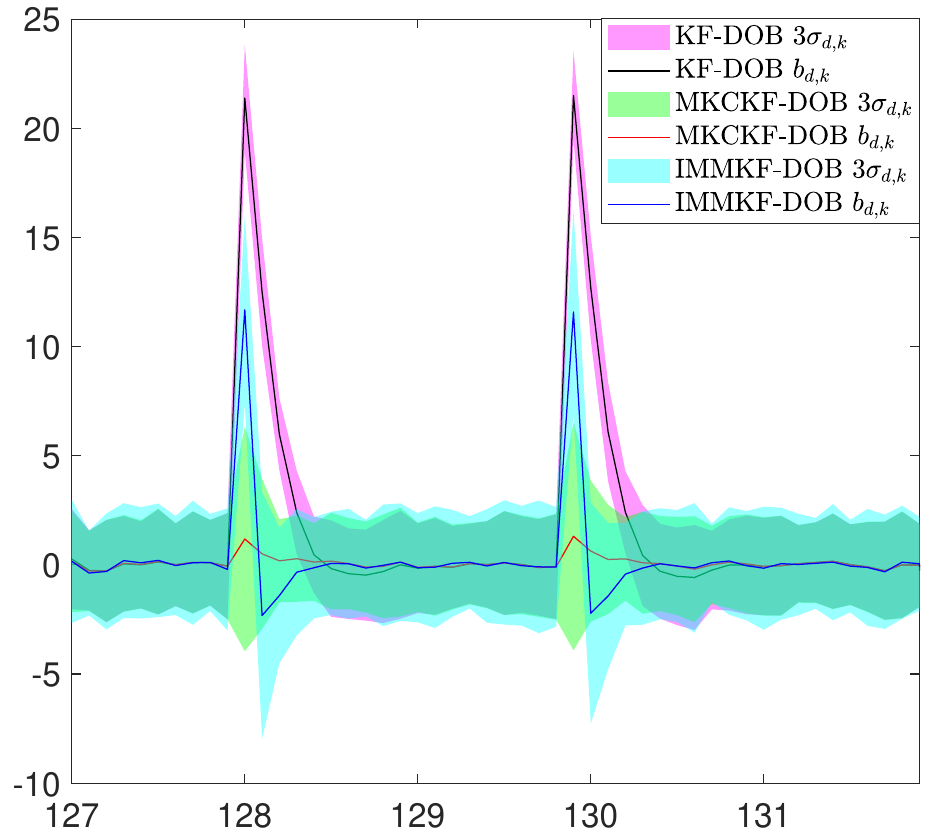}
				\label{kf-mkc-imm}
			\end{minipage}%
		}%
		\subfigure[]{
			\begin{minipage}[t]{0.45\linewidth}
				\centering
				\includegraphics[width=1.0\columnwidth]{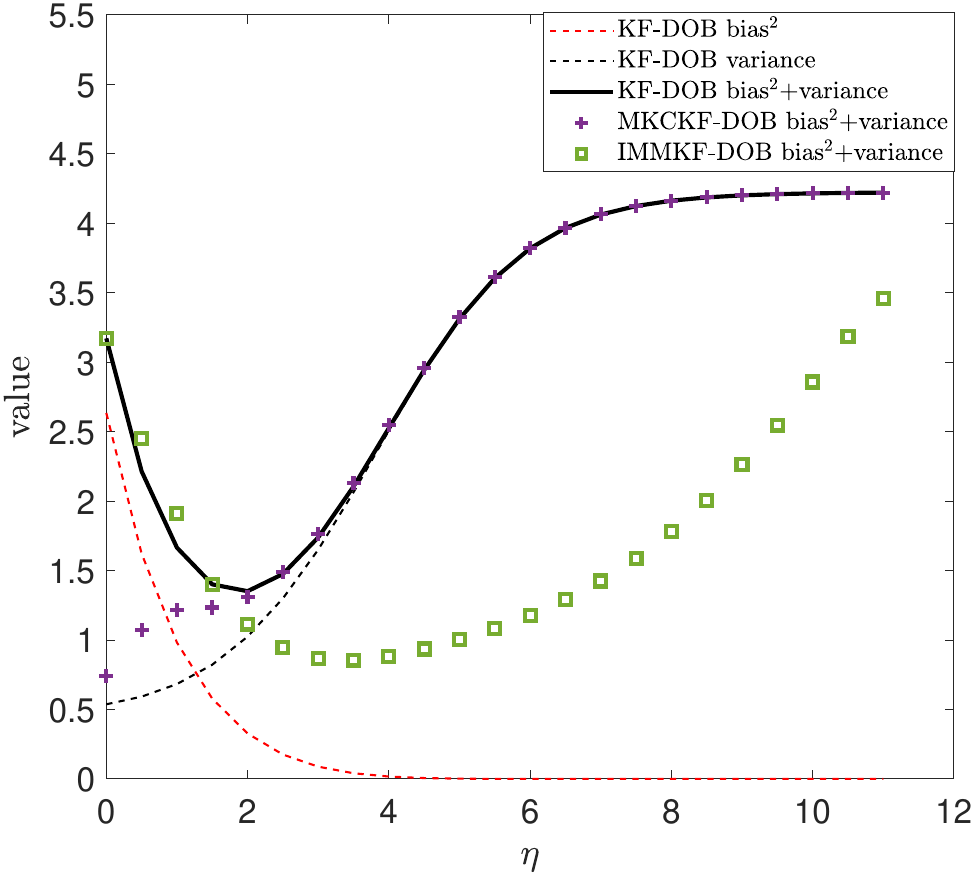}
				\label{biasvar}
			\end{minipage}%
		}%
		\caption{(a) The bias-variance visualization of KF-DOB, MKCKF-DOB, and IMMKF-DOB with 100 Monte Carlo runs. (b) Performance loss of different estimators with different $\eta$. Note that in MKCKF-DOB, the kernel bandwidth for the disturbance channel is set as $\varsigma_d=3+\eta$. In IMMKF-DOB, we use $D_1=D^{*}$ and $D_2=\exp(\eta)D^{*}$.}	
		\label{bias-variance}
	\end{figure}
	To visualize the bias-variance effects of the above two remedies, we conduct 100 Monte Carlo runs and visualize the corresponding results in Fig. \ref{kf-mkc-imm}. We observe that the $3\sigma_{d,k}$ region of the MKCKF and IMMKF is similar to that of the KF-DOB, but the bias effect is significantly mitigated. \textcolor{black}{To gain an intuitive understanding of the bias-variance effect in KF-DOB, we present a performance loss, which is a summation of the squared bias and variance, i.e., $\mathcal{L}=\bar{b}_d^2+\bar{\sigma}^2$ where $\bar{b}_d^2$ is the average squared bias and $\bar{\sigma}^2$ is the average variance within a time window (see Section \ref{tradeoff} for detailed definitions). Moreover, we use the same time window $t\in [126,132]$ as in Fig. \ref{bias_variance_tradeoff}. To provide a comprehensive comparison, we provide the corresponding performance loss of KF-DOB using different $D=\exp(\eta)D^{*}$, the loss of MKCKF-DOB with nominal disturbance covariance $D=D^{*}$ but with a different kernel bandwidth $\varsigma_d=3+\eta$, as well as the loss of IMMKF-DOB using the mixture of two models with $D_1=D^{*}$ and $D_2=\exp(\eta)D^{*}$ in Fig. \ref{biasvar} (note that they are all related with variable $\eta$). The results verify the effectiveness of the proposed remedies.}

	We summarize the root-mean-square-error performance of different estimators by conducting 100 Monte Carlo simulations in Table \ref{simr}. The program is executed in MATLAB on a laptop (Core(TM) i7-1360P, 2.2-GHz CPU, 16-GB RAM) and the time consumption of different algorithms is shown in the last column of Table \ref{simr}. We observe that the overall performances of MKCKF-DOB and IMMKF-DOB are better than the KF-DOB. We also see that the algorithm complexities of MKCKF-DOB and IMMKF-DOB are slightly higher than KF-DOB but are acceptable. 
	\textcolor{black}{
		\subsection{1-DOF Torsion System Example}
		We consider the 1-DOF torsion load system following \citet{zhao2016fusion}. It consists of a motor, a servo
		load, and a torsion load. The input of this system is the motor torque, while the disturbance is from unknown torsional loads. Using system parameters as shown in \citet{zhao2016fusion} and discretizing the continuous system using a zero-order hold approach with a sampling time $\mathrm{d}t=0.01$, one obtains
		\begin{equation}
			\begin{aligned}
				{x}_{k}&={F}_k {x}_{k-1}+{G}_{1,k} {u}_{k-1}+{G}_{2,k} {d}_{k-1}+{w}_{k}\\
				{y}_k&={H}_k {x}_{k}+{v}_{k}
				\label{linearfun}
			\end{aligned}
		\end{equation}
		where $x_k=[\theta_m,\theta_t,v_m,v_t]^{T} \in \mathbb{R}^{4}$ includes the angles at the motor side and the load side and the corresponding velocities, $y_k \in \mathbb{R}^{2}$ is the reading from encoders, $u_k=\tau_{m,k} \in \mathbb{R}$ is the motor torque and $d_k=\tau_{l,k} \in \mathbb{R}$ is caused by the unknown load. Moreover, one has   
		\begin{equation}\nonumber
			\begin{aligned}
				F_k&=\begin{bmatrix}
					0.9205 &   0.0795 &   0.0085 &   0.0003\\
					0.2045 &   0.7955 &   0.0007 &   0.0085\\
					-14.3468 &  14.3468 &   0.6872 &   0.0746\\
					37.5370 & -37.5370 &   0.1863 &   0.6405
				\end{bmatrix}\\
				G_{1,k}&=\begin{bmatrix}
					0.0826,
					0.0031,
					15.5568,
					1.2100
				\end{bmatrix}^{T}\\
				G_{2,k}&=\begin{bmatrix}
					0.0031,
					0.2076,
					1.2100,
					38.7470
				\end{bmatrix}^{T} \\
				H_k&=\textcolor{black}{\begin{bmatrix}
						1&0&0&0\\
						0&1&0&0
				\end{bmatrix}}.
			\end{aligned}
		\end{equation}
		To simultaneously estimate both the state and disturbance, by analogy to \eqref{sys}, we augment the disturbance and the state as a new state and construct the KF-DOB as follows:  
		\begin{equation}
			\begin{aligned}
				\mathrm{x}_k &= \Phi_k \mathrm{x}_{k-1} + \mathrm{G}_k \mathrm{u}_{k-1} + \mathrm{w}_k \\
				\mathrm{y}_k &= \mathrm{H}_k \mathrm{x}_k  + \mathrm{v}_k
			\end{aligned}
			\label{sys0}
		\end{equation}
		where 
		\begin{equation}\nonumber
			\scriptsize
			\begin{aligned}
				\mathrm{x}_{k}&=\begin{bmatrix}
					d_{k} \\
					x_{k}
				\end{bmatrix}, \quad \Phi_k=\begin{bmatrix}
					1   &0_{1 \times 3}\\
					G_{2,k} &F_k 
				\end{bmatrix},\\
				\mathrm{G}_k&=\begin{bmatrix}
					0\\
					G_{1,k}
				\end{bmatrix},\quad\mathrm{H}_k =\begin{bmatrix}
					{0}_{2 \times 1} & H_k
				\end{bmatrix}, \quad\mathrm{y}_k={y}_k.
			\end{aligned}
			\label{notdef}
	\end{equation}}
	
	\textcolor{black}{In the simulation, the process covariance has $Q=E(\mathrm{w}_k\mathrm{w}_k^{T})=\begin{bmatrix}
			D&0\\
			0&Q_x
		\end{bmatrix}$ and $Q_x=0.01 \mathbf{I}_{4}$. The measurement covariance has $R=E(\mathrm{v}_k\mathrm{v}_k^{T})=0.5 \mathbf{I}_{2}$. The ground truth disturbance is set to be
		\begin{equation}
			d_k=\begin{cases}
				10 + w_{d,k}, \quad  400 \le k < 600 \\
				w_{d,k}, \text{otherwise}
			\end{cases}
		\end{equation}
		where $w_d \sim \mathcal{N}(0,D^{*})$ and $D^{*}=0.01$. In KF-DOB, we use $D=D^{*}$, representing the nominal disturbance covariance. In MKCKF-DOB, we apply $\boldsymbol{\sigma}_p=[2,10^{8},10^{8},10^{8},10^{8}]^{T}$ and $\boldsymbol{\sigma}_r=[10^{8},10^{8}]^{T}$ according to Remark \ref{remark9}. In IMMKF-DOB, we use $\mathcal{P}=\begin{bmatrix}
			0.98&0.02\\
			0.5&0.5
		\end{bmatrix}$. Moreover, we use $D_1=\operatorname{exp}(0)D^{*}$ for Model 1 and $D_2=\operatorname{exp}(4)D^{*}$ for Model 2. The disturbance estimation performance of three estimators is shown in Fig. \ref{1dof_d}, where we find that their performances are similar when without disturbance jumps. However, two remedies, the MKCKF-DOB and IMMKF-DOB, have faster convergence speeds with abrupt disturbance changes. We further investigate the RMSE performance of KF-DOB when using $D=\exp(\eta) D^{*}$ where $\eta$ ranges from $0$ to $10$ and compare it with MKCKF-DOB and IMMKF-DOB in Fig. \ref{1dof_era}. We observe that the proposed methods are better than KF-DOB with any disturbance noise covariance. 
		\begin{figure}[htbp]
			\centering
			\subfigure[]{
				\begin{minipage}[t]{0.45\linewidth}
					\centering
					\includegraphics[width=1.0\columnwidth]{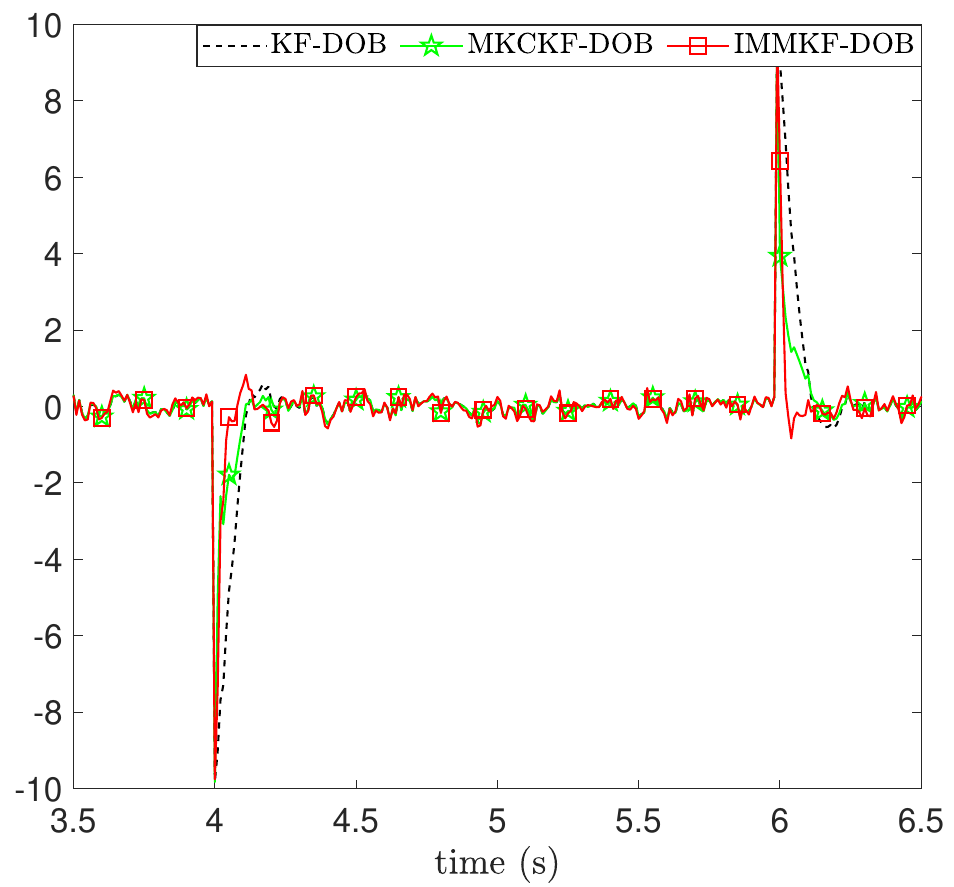}
					\label{1dof_d}
				\end{minipage}%
			}%
			\subfigure[]{
				\begin{minipage}[t]{0.451\linewidth}
					\centering
					\includegraphics[width=1.0\columnwidth]{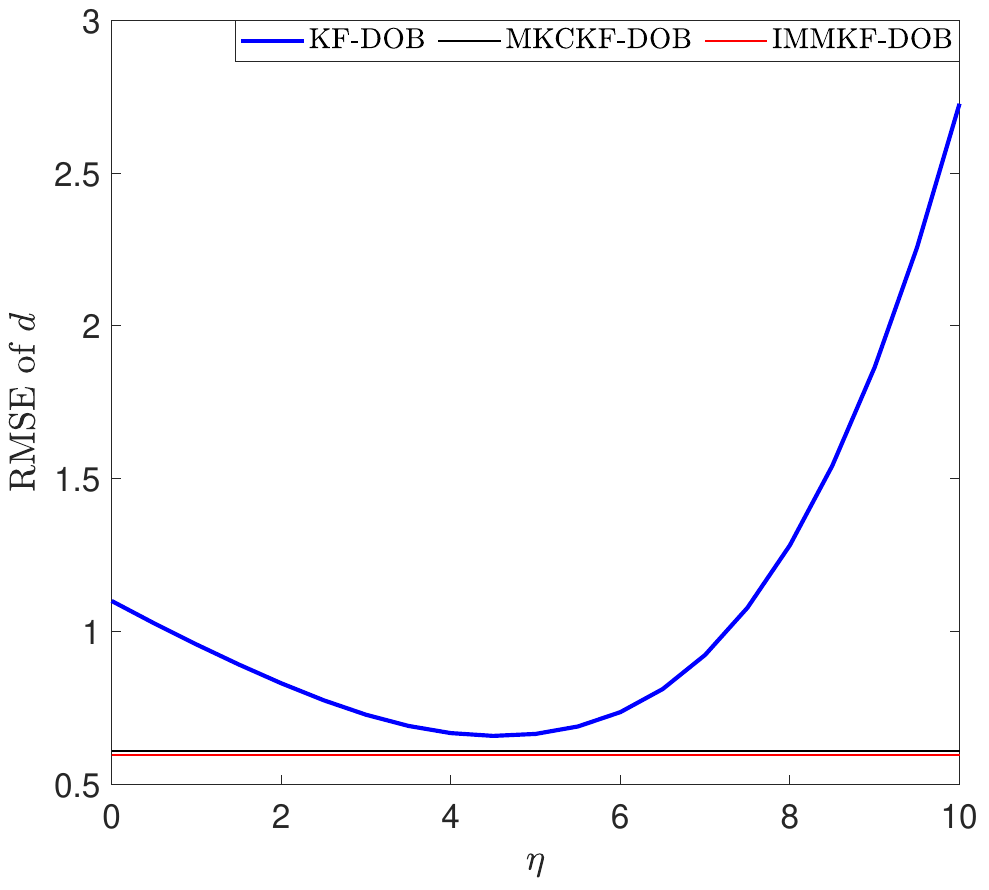}
					\label{1dof_era}
				\end{minipage}%
			}%
			\caption{\textcolor{black}{Disturbance estimation performances of KF-DOB, MKCKF-DOB, and IMMKF-DOB.}}
			\label{1doftorsion}	
		\end{figure}
	}
	
	\textcolor{black}{
		\subsection{1-DOF Manipulator Example}
		We consider a 1-DOF manipulator example according to \citet{b27}. The discrete state-space model of the 1-DOF manipulator has
		\begin{equation}
			\begin{aligned}
				\mathrm{x}_{k}&=\mathrm{\Phi}_k \mathrm{x}_{k-1}+\mathrm{G}_k \mathrm{u}_{k-1}+\mathrm{w}_{k}\\
				\mathrm{y}_{k}&=\mathrm{H}_k\mathrm{x}_{k}+\mathrm{v}_{k}
				\label{observer}
			\end{aligned}
		\end{equation}
		with
		\begin{equation}\nonumber
			\begin{aligned}
				\mathrm{\Phi}_k=\left[\begin{array}{ccc}
					1&0&0\\
					\frac{T}{{I}_{m}}& 1-\frac{b_{m}T}{{I}_{m}} &-\frac{k_{m}T}{{I}_{m}}\\
					0&T&1\\
				\end{array}\right]\\
				\mathrm{G}_k =\left[\begin{array}{c}
					0\\
					\frac{T}{{I}_{m}}\\
					0
				\end{array}\right], \mathrm{H}_k=\left[\begin{array}{l}
					0,0,1
				\end{array}\right]
			\end{aligned}
		\end{equation}
		where $\mathrm{u}_{k}=\bar{\tau}_{k}$ is the input torque, and $x_{k}=[d_{k}, \dot{\theta}_{k},\theta_{k}]^{\prime}$ including the disturbance, the angular velocity, and the angular acceleration, $T=0.01$ is the sampling time, $w_k=[w_{d,k},w_{x,k}]^{T}$ is the process noise, and $v_k$ is the measurement noise. In the simulation, we use ${I}_{m}=0.1$~Nm$\cdot$s$^2/\deg$, $b_m=1$~Nm$\cdot$s$/\deg$,  $k_m=0.1$~Nm$/\deg$. Moreover, we assume that the unknown disturbance has
		\begin{equation}\nonumber
			{d}_{k}=\left\{\begin{array}{l}
				50+{w}_{d,k},400\le k \le 600 \\
				{w}_{d,k}, \operatorname{otherwise}
			\end{array}\right..
		\end{equation}
		The desired angle follows $\theta_{d,k}=15 \sin (0.4\pi kT)$. To track the desired position, a PD controller with a feedforward term is used, referring to \cite{b27} for details.}
	
	\textcolor{black}{In KF-DOB and MKCKF-DOB, we use $Q=\begin{bmatrix}
			D^{*}&0\\
			0&Q_x
		\end{bmatrix}$ where $D^{*}=0.01$, $Q_x = 0.0001 \cdot \mathrm{I}_2$, and $R=0.0001$. Moreover, we apply $\mathbf{\sigma}_p=[1.5, 10^{8},10^{8},10^{8}]^{T}$ and $\mathbf{\sigma}_r=10^{8}$ in MKCKF-DOB. As for IMMKF-DOB, we use $\mathcal{P}=\begin{bmatrix}
			0.98&0.02\\
			0.5&0.5
		\end{bmatrix}$. Moreover, we use $D_1=\operatorname{exp}(0)D^{*}$ for Model 1 and $D_2=\operatorname{exp}(4)D^{*}$ for Model 2. The state error and tracking error of different estimators are shown in Figs. \ref{state_error} and \ref{track_error}, respectively. We observe that our proposed methods are significantly better than the conventional KF-DOB.
		\begin{figure}[htbp]
			\centering
			\subfigure[]{
				\begin{minipage}[t]{0.45\linewidth}
					\centering
					\includegraphics[width=1\columnwidth]{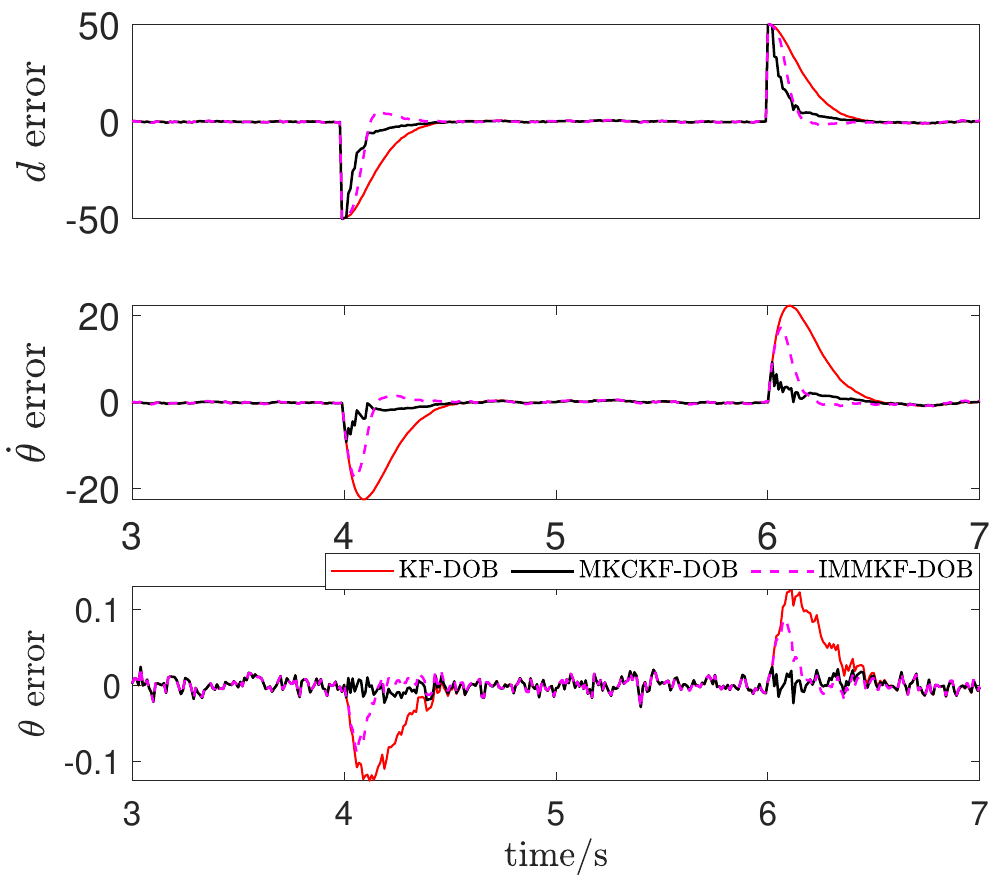}
					\label{state_error}
				\end{minipage}%
			}%
			\subfigure[]{
				\begin{minipage}[t]{0.5\linewidth}
					\centering
					\includegraphics[width=1\columnwidth]{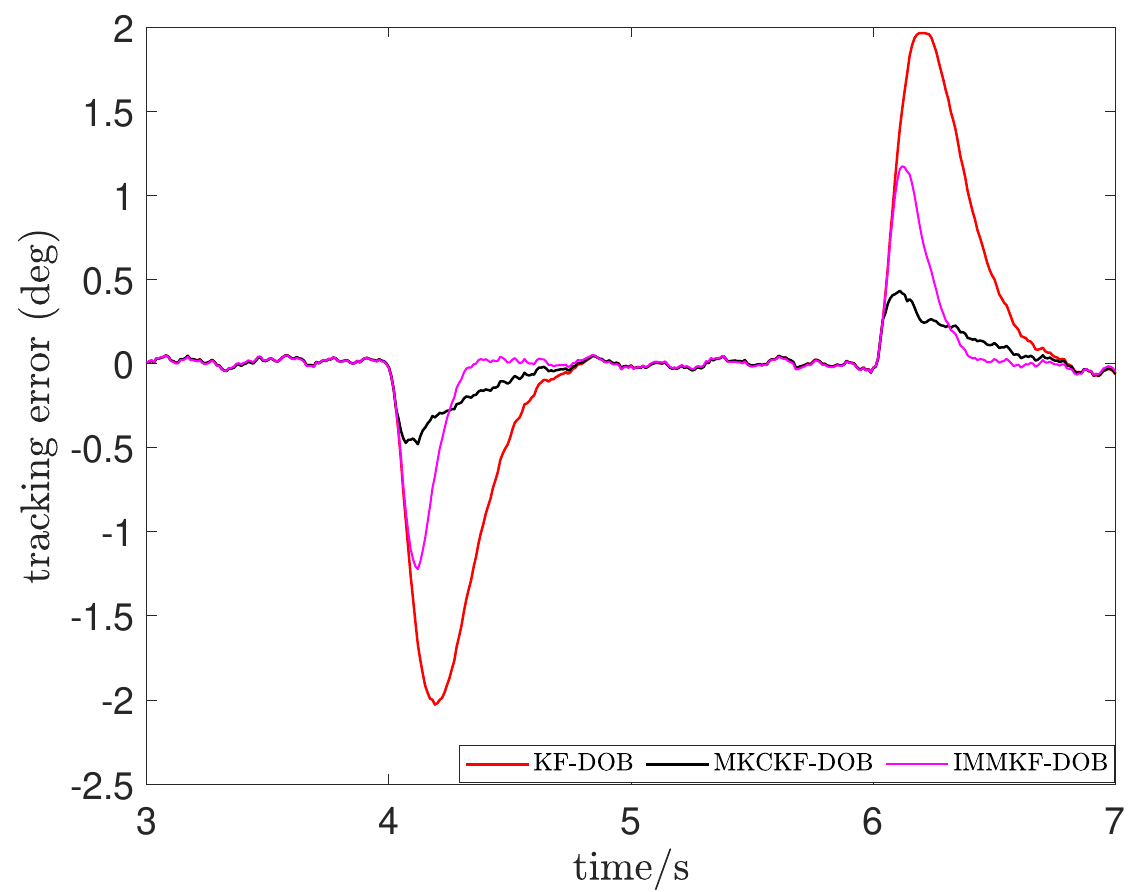}
					\label{track_error}
				\end{minipage}%
			}%
			\textcolor{black}{
				\caption{Estimation error and tracking error comparisons of the KF-DOB, MKCKF-DOB, and IMMKF-DOB.}
			}
	\end{figure}}
	\section{Conclusion}
	In the paper, we systematically investigate the bias-variance effects in KF-DOB and reveal that it is identical to the SISE estimator when applying infinite disturbance covariance. To meet the requirements of both timely and smooth disturbance estimates in practical applications, we propose two remedies: MKCKF-DOB and IMMKF-DOB. The superiority of the proposed methods is explained from the perspective of ``adaptive'' or ``switched'' disturbance covariance and is verified in extensive simulations and experiments. It is worth noting that although MKCKF-DOB and IMMKF-DOB have better performances compared with the existing approaches, MKCKF-DOB needs to tune the kernel bandwidths and IMMKF-DOB needs to tune the Markov transition probability matrix empirically. In the future, we will devote ourselves to developing adaptive mechanisms to avoid these time-consuming procedures.
	\section{Appendix}
		\subsection{Proof of Lemma \ref{lemma5}}
	\label{appendix4}
	According to the posterior error covariance update equation and its corresponding information form update formula as shown in \eqref{infp}, one has
	\begin{equation}
		\begin{aligned}
			\mathrm{P}_{k|k}&=(\mathrm{I}-\mathrm{K}_k \mathrm{H}_k )\mathrm{P}_{k|k-1} \\
			&=\Big(\mathrm{P}_{k|k-1}^{-1} +\mathrm{H}_k^{T} \mathrm{R}_k^{-1} \mathrm{H}_k \Big)^{-1}	\\
			&=\Big(\mathrm{P}_{k|k-1}^{-1} +\mathrm{P}_{k|k-1}^{-1} \mathrm{P}_{k|k-1} \mathrm{H}_k^{T} \mathrm{R}_k^{-1} \mathrm{H}_k \Big)^{-1}\\
			&=\Big(\mathrm{P}_{k|k-1}^{-1} \big(\mathrm{I}+ \mathrm{P}_{k|k-1} \mathrm{H}_k^{T} \mathrm{R}_k^{-1} \mathrm{H}_k \big)\Big)^{-1}\\
			&=\big(\mathrm{I}+ \mathrm{P}_{k|k-1} \mathrm{H}_k^{T} \mathrm{R}_k^{-1} \mathrm{H}_k \big)^{-1}\mathrm{P}_{k|k-1}\\
		\end{aligned}
		\label{identity2}
	\end{equation}
	Then, according to the first and final line of \eqref{identity2}, one obtains \eqref{ident}. This completes the proof.
	\subsection{Derivation of Equation \eqref{batkf}}
	\label{appendix1}
	Based on the unbiasedness constraint, one has 
	\begin{equation}
		E(\mathrm{\hat{x}}_k)=E(\mathrm{x}_k).
		\label{ubiased}
	\end{equation}
	Applying \eqref{state1} and \eqref{batest2} into \eqref{ubiased} with $\mathrm{Y}_{1,k}$ specified in \eqref{vecsys}, on arrives
	\begin{equation}
		\mathcal{H}_{1, k}^{{s}}= \mathrm{\phi}_{k}^{1}-\mathcal{H}_{1, k}^{{h}}\mathrm{H}_{1,k}.
		\label{unbias}
	\end{equation}
	According to orthogonality principle, the estimator is optimal if $\mathrm{e}_k \triangleq \mathrm{x}_k-\mathrm{\hat{x}}_k$ is orthogonal to $\mathrm{Y}_{1,k}$ and $\mathrm{x}_0$, i.e.,
	\begin{equation}
		E\big((\mathrm{x}_k- \mathcal{H}_{1, k}^{{h}} \mathrm{Y}_{1,k} - \mathcal{H}_{1, k}^{{s}} \mathrm{x}_{0})\mathrm{Y}_{1,k}^{T}\big)=0.
		\label{orth}
	\end{equation}
	Substituting the expression of $\mathrm{Y}_{1,k}$ as shown \eqref{vecsys} and concerning that $\mathrm{x}_0$, $\mathrm{W}_{1,k}$, and $\mathrm{V}_{1,k}$ are mutually independent, \eqref{orth} can be rewritten as
	\begin{equation}
		\mathcal{B}_{1,k} \bar{P}_0 \mathrm{H}_{1,k}^{T} + 	\mathcal{W}_{1,k}\mathrm{Q}_{1,k}\mathrm{D}_{1,k}^{T}+	\mathcal{V}_{1,k}\mathrm{R}_{1,k}=0
		\label{batest}
	\end{equation} 
	where 
	\begin{equation}
		\begin{aligned}
			\mathcal{B}_{1,k}&=\phi_{k}^{1}-\mathcal{H}_{1, k}^{{h}} \mathrm{H}_{1,k}- \mathcal{H}_{1, k}^{{s}} \\
			\mathcal{W}_{1,k}&=\mathrm{G}_{1,k}^{rk}-\mathcal{H}_{1, k}^{{h}}\mathrm{D}_{1,k},~\mathcal{V}_{1,k}=\mathcal{H}_{1, k}^{{h}}\\
		\end{aligned}
	\end{equation}
	and $\mathrm{\bar{P}}_0=E(\mathrm{x}_0 \mathrm{x}_0^{T})$, $\mathrm{Q}_{1,k} = E(\mathrm{W}_{1,k} \mathrm{W}_{1,k}^{T})$, $\mathrm{R}_{1,k} = E(\mathrm{V}_{1,k} \mathrm{V}_{1,k}^{T})$. Since $\mathrm{x}_0$ is known, the expectation operator in $\mathrm{\bar{P}}_0$ can be removed, i.e., $\mathrm{\bar{P}}_0=E(\mathrm{x}_0 \mathrm{x}_0^{T})=0$. By collecting the terms in \eqref{batest}, one obtains the expression of $\mathcal{H}_{1, k}^{{h}}$ as shown in \eqref{gainy}. Finally, according to \eqref{batest2} and \eqref{unbias}, one arrives \eqref{batkf}.
	\subsection{Derivation of Equation \eqref{batkf2}}
	\label{appendix3}
	According to the unbiasedness condition $E(\mathrm{x}_k) = E(\mathrm{\hat{x}}_k)$ where $\mathrm{\hat{x}}_k$ is shown in \eqref{batest4} and $\mathrm{x}_k$ is shown in \eqref{state1} with $\mathrm{Y}_{1,k}$ specified in \eqref{vecsys}, one obtains
	\begin{equation}
		\mathcal{\bar{H}}_{1, k}^{{s}}= \mathrm{\phi}_{k}^{1}-\mathcal{\bar{H}}_{1, k}^{{h}}\mathrm{H}_{1,k}.
		\label{unbias1}
	\end{equation}
	According to the orthogonality principle, one has $E[(\mathrm{x}_k-\mathrm{\hat{x}}_k)\mathrm{Y}_{1,k}^{T}]=0$. In a similar manner to the obtainment of \eqref{gainy}, one can obtain \eqref{gainy2}. Finally, substituting the expression of \eqref{unbias1} into \eqref{batkf2} completes the proof.
	\subsection{Proof of Proposition \ref{prop2}}
	\label{addendixprop}
	One can decompose the results of $\mathrm{\hat{x}_k}$ into two components: the first component is driven by $\mathrm{{x}}_0=0$ and $\mathrm{P_0}=\mathrm{\bar{P}_0}$ with normal $\mathrm{y}_k$, while the second is driven by $\mathrm{{x}}_0 =\bar{x}_0$ and $\mathrm{P_0}=\mathrm{\bar{P}_0}$ but $\mathrm{y}_k=0$ for all $k$. In the first case, \eqref{batkf2} reduces to $\mathrm{\hat{x}}_k=\mathrm{\hat{x}}_k^{\bar{h}}$. In Kalman filtering, this gives \eqref{equ11}. In the second case, \eqref{batkf2} reduces to $\mathrm{\hat{x}}_k=\mathrm{\hat{x}}_k^{\bar{s}}$. In Kalman filtering, this gives \eqref{equ22} starting from  $\mathrm{\hat{x}}_0^{s}=\mathrm{\bar{x}_0}$. By superimposing \eqref{equ11} and \eqref{equ22}, one can recover the KF as shown in \eqref{pk} with an initial guess $\mathrm{x}_0 \sim \mathcal{N}(\mathrm{\bar{x}_0},\mathrm{\bar{P}_0})$, which completes the proof.
	\subsection{Proof of Lemma \ref{lemmazz}}
	\label{prooflemmazz}
	First, we prove the first part of Lemma \ref{lemmazz}. Let $y= (\mathrm{X}_k+\mathrm{Y}_k)^{T} x$ where $x$ is an arbitrary vector. According to $\|\mathbf{Z}_{+}\| = \|\mathbf{Z}_{+}^{T}\|\le 1$, we have 
	$$
	\|\mathbf{Z}_{+}^{T} y \| \le \|\mathbf{Z}_{+} \| \| y \| \le \| y \|.
	$$
	It implies that $\|y\|^{2} \ge \|\mathbf{Z}_{+}^{T} y \|^{2}$. Substituting the expression of $y$ and $\mathbf{Z}_{+}$ gives
	\begin{equation}
		x^{T}(\mathrm{X}_k+\mathrm{Y}_k)(\mathrm{X}_k+\mathrm{Y}_k)^{T}x \succeq x^{T}\mathrm{X}_k \mathrm{X}_k^{T}x .
	\end{equation}
	We then focus on the second part of Lemma \ref{lemmazz}. Let $y= \mathrm{X}_k^{T} x$ where $x$ is an arbitrary vector. According to $\|\mathbf{Z}_{-}\| = \|\mathbf{Z}_{-}^{T}\|\le 1$, we have $ \|\mathbf{Z}_{-}^{T} y \| \le \|\mathbf{Z}_{-} \| \| y \| \le \| y \|$. It implies that $\|y\|^{2} \ge \|\mathbf{Z}_{-}^{T} y \|^{2}$.  
	Substituting the expression of $y$ and $\mathbf{Z}_{-}$ gives
	\begin{equation}
		\begin{aligned}
			x^{T}\mathrm{X}_k\mathrm{X}_k^{T}x  \succeq x^{T}(\mathrm{X}_k+\mathrm{Y}_k)(\mathrm{X}_k+\mathrm{Y}_k)^{T}x
		\end{aligned}
	\end{equation}
	This completes the proof.
	\subsection{Proof of Theorem \ref{theorem1}}
	\label{appendix5}
	Denote the discrete-time algebraic Riccati equation as 
	\begin{equation}\nonumber
		g(X)=\Phi_k X \Phi_k^{T} + \mathrm{Q}_k  - \Phi_k X \mathrm{H}_k^{T} (\mathrm{H}_k X \mathrm{H}_k^{T}+\mathrm{R}_k)^{-1} \mathrm{H}_k  X \Phi_k^{T}.
	\end{equation}
	It follows that $\mathrm{P}_{k+1|k}=g(\mathrm{P}_{k|k-1})$. According to \citet{c14}, $g(X)\succeq g(Y)$ if $X\succeq Y$ and $\mathrm{Q}_k^{X} \succeq \mathrm{Q}_k^{Y}$ where $\mathrm{Q}_k^{X}$ and $\mathrm{Q}_k^{Y}$ are the corresponding process covariance matrices. Accordingly, it follows that $\mathrm{P}_{k|k-1}^{o} \preceq \mathrm{P}_{k|k-1}^{u}$ for $\Delta \mathrm{Q} \succeq 0$ and  $\mathrm{P}_{k|k-1}^{o} \succeq \mathrm{P}_{k|k-1}^{u}$ for $\Delta \mathrm{Q} \preceq 0$ for $k \ge 1$. 
	Based on Lemma \ref{lemma5}, it follows that
	\begin{equation}
		\begin{aligned}
			\mathrm{M}_k^{o}&=\big(\mathrm{I}+ \mathrm{P}_{k|k-1}^{o} \mathrm{H}_k^{T} \mathrm{R}_k^{-1} \mathrm{H}_k \big)^{-1}\\
			\mathrm{M}_k^{u}&=\big(\mathrm{I}+ \mathrm{P}_{k|k-1}^{u} \mathrm{H}_k^{T} \mathrm{R}_k^{-1} \mathrm{H}_k \big)^{-1}\\
		\end{aligned}
	\end{equation}
	Based on \eqref{defXY}, one has
	\begin{equation}
		\begin{aligned}
			&\Big(\mathrm{(\mathrm{M}_k^{u})^{T}\mathrm{M}_k^{u}}\Big)^{-1}=(\mathrm{X}_k+\mathrm{Y}_k)(\mathrm{X}_k+\mathrm{Y}_k)^{T}\\
			&=\Big(\mathrm{(\mathrm{M}_k^{o})^{T}\mathrm{M}_k^{o}}\Big)^{-1}+\mathrm{X}_k \mathrm{Y}_k^{T}+\mathrm{Y}_k \mathrm{X}_k^{T}+ \mathrm{Y}_k\mathrm{Y}_k^{T}
		\end{aligned}
	\end{equation}
	According to Assumption \ref{assump1}, it follows that $(\mathrm{M}_k^{u})^{T}\mathrm{M}_k^{u} \preceq (\mathrm{M}_k^{o})^{T}\mathrm{M}_k^{o}$ if $\Delta \mathrm{Q} \succeq \mathrm{Q}_{\epsilon+}$ and $(\mathrm{M}_k^{u})^{T}\mathrm{M}_k^{u} \succeq (\mathrm{M}_k^{o})^{T}\mathrm{M}_k^{o}$ if $\Delta \mathrm{Q} \preceq \mathrm{Q}_{\epsilon-}$. Subsequently, by denoting $z\triangleq {\Phi}_k  \mathrm{\hat{x}}_{k-1}^{b}$, it follows that
	$C_{\gamma,k}^{o}=z^{T}(\mathrm{M}_k^{o})^{T}\mathrm{M}_k^{o}z \ge z^{T}(\mathrm{M}_k^{u})^{T}\mathrm{M}_k^{u}z=C_{\gamma,k}^{u}$ if $\Delta \mathrm{Q} \succeq \mathrm{Q}_{\epsilon+}$ and $C_{\gamma,k}^{o}=z^{T}(\mathrm{M}_k^{o})^{T}\mathrm{M}_k^{o}z \le z^{T}(\mathrm{M}_k^{u})^{T}\mathrm{M}_k^{u}z=C_{\gamma,k}^{u}$ if $\Delta \mathrm{Q} \preceq \mathrm{Q}_{\epsilon-}$. This completes the proof.
	\subsection{Proof of Proposition \ref{prop3}}
	\label{appendix6}
	According to \eqref{qsys}, it is clear that $\mathrm{Q}_k^{u}\to \infty$ when $\Delta \mathrm{Q} \to \infty$. Then, one obtains $\mathrm{M}_k^{u}\to 0$ according to \eqref{ident} under the premise that $\mathrm{H}_k^{T} \mathrm{R}_k^{-1} \mathrm{H}_k$ is a PD matrix. Consequently, one arrives
	\begin{equation}
		\lim_{\Delta \mathrm{Q} \to \infty}	C_{\gamma,1}^{u}=(\mathrm{\hat{x}}_{0}^{b})^{T} \Phi_1^{T}(\mathrm{M}_k^{u})^{T} \mathrm{M}_k^{u} \Phi_1 \mathrm{\hat{x}}_{0}^{b} \to 0
	\end{equation}
	which implies that the estimation bias $\mathrm{\hat{x}}_k^{b}$ convergences to zero after receiving a single measurement if $\Delta \mathrm{Q} \to \infty$.
	\subsection{Proof of Theorem \ref{theorem2}}
	\label{appendix7}
	According to \eqref{infp} and \eqref{infpf}, one obtains 
	\begin{equation}
		\begin{aligned}
			(\mathrm{P}_{k|k}^{f})^{-1} - (\mathrm{P}_{k|k})^{-1} &= (\Phi_k \mathrm{P}_{k-1|k-1} \Phi_k^{T}  +  \mathrm{Q}_k^{u})^{-1} \\
			&- (\Phi_k \mathrm{P}_{k-1|k-1} \Phi_k^{T}  +  \mathrm{Q}_k)^{-1}.
		\end{aligned}
	\end{equation}
	It follows that $\mathrm{P}_{k|k} \preceq\mathrm{P}_{k|k}^{f1} \preceq \mathrm{P}_{k|k}^{f2}$ if $\mathbf{0} \preceq \Delta \mathrm{Q}_1 \preceq \Delta \mathrm{Q}_2$ and $\mathrm{P}_{k|k}\succeq \mathrm{P}_{k|k}^{f1} \succeq \mathrm{P}_{k|k}^{f2}$ if $\mathbf{0} \succeq \Delta \mathrm{Q}_1 \succeq \Delta \mathrm{Q}_2 \succeq - \mathrm{Q}_k$. 
	
	Then, we compare $\mathrm{P}_{k|k}^{t1}$ and $\mathrm{P}_{k|k}^{t2}$.  According to \eqref{pk} and \eqref{pkt}, one has
	\begin{equation}
		\begin{aligned}
			\mathrm{P}_{k \mid k}^{t}-\mathrm{P}_{k \mid k}= & \mathrm{C}_k \mathrm{R}_k   \times\left(\mathrm{A}_k+\mathrm{A}_k \mathrm{B}_k^{-1} \mathrm{A}_k \right)^{-1} \\
			& \times \mathrm{A}_k\left(\left(\mathrm{A}_k+\mathrm{A}_k \mathrm{B}_k^{-1} \mathrm{A}_k \right)^{-1}\right)^{T}  \left(\mathrm{R}_k \right)^T \mathrm{C}_k^T
			\label{ptrue}
		\end{aligned}
	\end{equation}
	where
	\begin{equation}
		\begin{aligned}
			& \mathrm{A}_k=\mathrm{H}_k \mathrm{P}_{k \mid k-1} \mathrm{H}_k^T+\mathrm{R}_k \\
			&\mathrm{B}_k = \mathrm{H}_k \Delta \mathrm{Q} \mathrm{H}_k^T,~ \mathrm{C}_k=\mathrm{H}_k^T\left(\mathrm{H}_k \mathrm{H}_k^T\right)^{-1}.
			\label{ABC}
		\end{aligned}
	\end{equation}
	The above equation indicates that $\mathrm{P}_{k \mid k}^{t} \succeq 	\mathrm{P}_{k \mid k}$ always holds due to its quadratic form. Since $\Delta \mathrm{Q}$ only appears in $\mathrm{B}_k$, one has $\mathrm{P}_{k|k} \preceq \mathrm{P}_{k|k}^{t1} \preceq \mathrm{P}_{k|k}^{t2}$ if $\mathbf{0} \preceq \Delta \mathrm{Q}_1 \preceq \Delta \mathrm{Q}_2$ and $\mathrm{P}_{k|k} \preceq \mathrm{P}_{k|k}^{t1} \preceq \mathrm{P}_{k|k}^{t2}$ if $\mathbf{0} \succeq \Delta \mathrm{Q}_1 \succeq \Delta \mathrm{Q}_2 \succeq - \mathrm{Q}_k$. This completes the proof.
	\subsection{Proof of Theorem \ref{theorem3}}
	\label{proofth3}
	Since KF is optimal at its steady state in the minimum mean-square-error sense when applying the correct process and measurement covariance matrices, we conclude that  $\mathrm{P}_{\infty}^{t,dob}$ is minimized when applying $\Delta \mathrm{Q}=\textbf{0}$. At the same time, according to  Theorem \ref{theorem1}, one can deduce that $\Delta \mathrm{Q}=\textbf{0}$ yields the slowest bias convergence under Assumption \ref{assump2} for any $\Delta \mathrm{Q}_{d} \succeq \textbf{0}$. Meanwhile, according to Proposition \ref{prop3} and Theorem \ref{theorem2}, one can conclude that the true error covariance is maximized but the bias convergence rate is fastest (i.e., infinite convergence rate) by applying $\Delta \mathrm{Q}_{d} \to \infty$.
	\subsection{Proof of Theorem \ref{theorem4}}
	\label{proofth4}
	We first show that KF-DOB is unbiased if and only if $\Delta \mathrm{Q}_{d} \to \infty$. According to Assumption \ref{kfdobcov}, we know that KF-DOB is unbiased at time step $j$ before its next disturbance jump, i.e., $\mathrm{\hat{x}}_j=E(\mathrm{x}_j)$. According to partitioned matrix inversion lemma \citep{c15_1}, if matrices $A$ and $D$ are invertible, one has
	\begin{equation}
		\begin{bmatrix}
			A&B\\
			C&D
		\end{bmatrix}^{-1}=\begin{bmatrix}
			A^{-1}+A^{-1} B E C A^{-1}&-A^{-1} B E\\
			-E C A^{-1}&E
		\end{bmatrix}
		\label{block1}
	\end{equation}
	where
	$
	E = \left(D-C A^{-1} B\right)^{-1}.
	$
	Based on the bias propagation equation \eqref{biaseffect} and the ``initialization error" shown in \eqref{inierror}, one obtains 
	\begin{equation}
		\bar{\mathrm{x}}_{j+1}^{\mathrm{b,dob}}=\bar{\Phi}_k\bar{\mathrm{x}}_j^{\mathrm{b,dob}}
		\label{biasj1}
	\end{equation}
	where $\bar{\Phi}_k=\Phi_k \mathrm{M}_k^{u}$ and $\mathrm{M}_k^{u}=\big(\mathrm{I}+ \mathrm{P}_{k|k-1}^{u} \mathrm{H}_k^{T} \mathrm{R}_k^{-1} \mathrm{H}_k \big)^{-1}$ (see Lemma \ref{lemma5}). Denote $(\mathrm{M}_k^{u})^{-1}=S=\begin{bmatrix}
		S_{11}&S_{12}\\
		S_{21}&S_{22}
	\end{bmatrix}$ where $S_{11}$ and $S_{22}$ are square matrices with the same dimension of $\Delta \mathrm{Q}_{d}$ and $\Delta \mathrm{Q}_{x}$. As
	$\Delta \mathrm{Q}_d \to \infty$, according to the error covariance prediction formula $\mathrm{P}_{k|k-1}^{u}= \Phi_k \mathrm{P}_{k-1|k-1}^{u} \Phi_k^{T}  +  \mathrm{Q}_k+\begin{bmatrix}
		\Delta \mathrm{Q}_d&0\\
		0&\Delta \mathrm{Q}_x
	\end{bmatrix}$, one has $S_{11} \to \infty$ and $S_{11}^{-1}\to 0$. It follows that
	\begin{equation}
		\mathrm{M}_k^{u}=\begin{bmatrix}
			0&0\\
			0&S_{22}^{-1}
		\end{bmatrix}
		\label{mu}
	\end{equation}
	according to \eqref{block1}. Subsequently, one has $\bar{\mathrm{x}}_{j+1}^{\mathrm{b,dob}}=0$ by substituting the expressions of $\mathrm{M}_k^{u}$ in \eqref{mu} and $\bar{\mathrm{x}}_{j}^{\mathrm{b,dob}}$ in \eqref{inierror} into \eqref{biasj1}. It indicates that $\mathrm{\hat{x}}_{j+1}=E(\mathrm{x}_{j+1})$ and KF-DOB is unbiased. Conversely, KF-DOB is always biased  since $\bar{\mathrm{x}}_{j+1}^{\mathrm{b,dob}} \neq 0$ if $\Delta \mathrm{Q}_{d} \nrightarrow \infty$. These facts reveal that KF-DOB is unbiased if and only if $\mathrm{Q}_{d} \to \infty$.
	
	We then prove that $\Delta \mathrm{Q}_{x}=\textbf{0}$ gives the minimum variance estimator among all unbiased estimators. According to Lemma \ref{lemma6} and Theorem \ref{theorem2}, one can observe that any additional perturbation on the process covariance will inflate the true error covariance. Hence, one can infer that $\Delta \mathrm{Q}_x=0$ gives a minimum variance estimator among all $\Delta \mathrm{Q}_x$. This completes the proof. 
	\subsection{Proof of Theorem \ref{theorem5}}
	\label{pftheorem5}
	Before proceeding, we make the following assumptions: $P_{k-1|k-1}^{dd} \to D_k$ and $P_{k-1|k-1}^{dx}$ is negligible compared with $D_k G_k^{T}$ as $D_k \to \infty$. Under these assumptions, according to the standard Kalman filter equation, it follows that 
	\begin{equation}
		\scriptsize
		\begin{aligned}
			P_{k|k-1}^{dd} \to & 2D,~P_{k|k-1}^{dx} \to  DG_k^{T}\\
			P_{k|k-1}^{xx} \to & G_k D_k G_k^{T} +F_k P^{xx} F_k ^{T}+ Q_k\\
			K_k^{x}=& P_{k|k-1}^{xx}H_k^{T} (H_k P_{k|k-1}^{xx} H_k^{T}+R_k)^{-1}\\
			M_{k}^{d}\to& D_k G_k^{T} H_k^{T}(H_k P_{k|k-1}^{xx} H_k^{T}+R_k)^{-1}\\
			P_{k|k}^{xx}& =(I-K_{k}H_k)P_{k|k-1}^{xx}\\
			P_{k|k}^{dd}&\to 2D_k- D_k G_k^{T} H_k^{T}(H_k P_{k|k-1}^{xx} H_k^{T}+R_k)^{-1} \\
			&\times H_k  G_k D_k \to D_k \\
			x_{k|k}=& F_k x_{k-1|k-1} + G_k d_{k-1|k-1}+ K_k^{x}\\
			&(y_k-H_k F_k x_{k-1|k-1}- H_k G_k d_{k-1|k-1})\\
			d_{k|k} =&	d_{k-1|k-1}+  M_{k}^{d}(y_k-H_k F_k x_{k-1|k-1}-H_k G_k d_{k-1|k-1})
		\end{aligned}
		\label{akfdob}
	\end{equation}
	One can observe that a necessary condition for the identity of the above equation with estimator \eqref{nkfdob} is that the following equations hold  as $D_k \to \infty$, 
	\begin{subequations}
		\begin{align}
			d_{k-1|k-1} - M_{k}^{d} H_k G_k d_{k-1|k-1} &\to 0  \label{sub1} \\
			G_k d_{k-1|k-1} - K_{k}^{x} H_k G_k d_{k-1|k-1} &\to 0 \label{sub2}
		\end{align}
	\end{subequations}
	By observation, one obtains $M_{k}^{d} \to M_{k}$ and $K_k^{x} \to K_k$ as $D_k \to \infty$ by comparing \eqref{akfdob} and \eqref{nkfdob}. Then, according to Lemma \ref{lemma9}, it follows that $M_{k}H_kG_k=M_{k}^{d}H_kG_k=I$ and hence \eqref{sub1} holds. Similarly, based on Lemma \ref{lemma9}, as $D_k \to \infty$, one obtains $K_k^{x} H_k G_k = K_k H_k G_k = G_k M_k^{*}H_k G_k + K_k^{*}(H_k G_k-H_k G_k M_k^{*}H_k G_k)=G_k$, hence \eqref{sub2} holds. A remaining issue is to prove $P_{k-1|k-1}^{dd} \to D_k$ and $P_{k-1|k-1}^{dx} \ll D_k G^{T}$ as $D_k \to \infty$. Without loss of generality, one can set the initial covariance as $P_{0|0}=\begin{bmatrix}
		D_k& 0\\
		0& P_{0|0}^{x}
	\end{bmatrix}$. As $D_k \to \infty$, according to the propagation of $P_{k-1|k-1}^{dd}$ and $P_{k-1|k-1}^{dx}$ as shown in \eqref{kfdob}, one can deduce that $P_{k-1|k-1}^{dd} \to D_k$ and $P_{k-1|k-1}^{dx}$ is negligible compared with $D_k G_k^{T}$ for $k\ge 2$. This completes the proof.
	\bibliographystyle{plainnat}
	\bibliography{reference_dob} 
\end{document}